\documentclass[twocolumn]{aastex61}
\usepackage{multirow}
\newcommand{\numax}{\mbox{$\nu_{\rm max}$}}
\newcommand{\Dnu}{\mbox{$\Delta \nu$}}

\newcommand{\muHz}{\mbox{$\mu$Hz}}
\newcommand{\kep}{\mbox{\textit{Kepler}}}
\newcommand{\teff}{\mbox{$T_{\rm eff}$}}
\newcommand{\logg}{\mbox{$\log g$}}
\newcommand{\feh}{\mbox{$\rm{[Fe/H]}$}}
\newcommand{\msun}{\mbox{M$_{\sun}$}}
\newcommand{\rsun}{\mbox{R$_{\sun}$}}

\shorttitle{Ensemble asteroseismology of 16,000 \kep\ oscillating red giants}
\shortauthors{Yu et al.}

\begin{document}
\title{Asteroseismology of 16,000 \kep\ red giants: Global oscillation parameters, Masses, and radii}

\correspondingauthor{Jie Yu}
\email{jiyu9229@uni.sydney.edu.au}

\author[0000-0002-0007-6211]{Jie Yu}
\affiliation{Sydney Institute for Astronomy (SIfA), School of Physics, University of 
Sydney, NSW 2006, Australia}
\affiliation{Stellar Astrophysics Centre, Department of Physics and Astronomy, Aarhus 
University, Ny Munkegade 120, DK-8000 Aarhus C, Denmark}

\author{Daniel Huber}
\affiliation{Institute for Astronomy, University of Hawai`i, 2680 Woodlawn Drive, Honolulu, HI 96822, USA}
\affiliation{Sydney Institute for Astronomy (SIfA), School of Physics, University of Sydney, NSW 2006, Australia}
\affiliation{SETI Institute, 189 Bernardo Avenue, Mountain View, CA 94043, USA}
\affiliation{Stellar Astrophysics Centre, Department of Physics and Astronomy, Aarhus 
University, Ny Munkegade 120, DK-8000 Aarhus C, Denmark}

\author{Timothy R.\ Bedding}
\affiliation{Sydney Institute for Astronomy (SIfA), School of Physics, University of 
Sydney, NSW 2006, Australia}
\affiliation{Stellar Astrophysics Centre, Department of Physics and Astronomy, Aarhus 
University, Ny Munkegade 120, DK-8000 Aarhus C, Denmark}

\author{Dennis Stello}
\affiliation{School of Physics, University of New South Wales, NSW 2052, Australia}
\affiliation{Sydney Institute for Astronomy (SIfA), School of Physics, University of 
Sydney, NSW 2006, Australia}
\affiliation{Stellar Astrophysics Centre, Department of Physics and Astronomy, Aarhus 
University, Ny Munkegade 120, DK-8000 Aarhus C, Denmark}

\author{Marc Hon}
\affiliation{School of Physics, University of New South Wales, NSW 2052, Australia}

\author{Simon J.\ Murphy}
\affiliation{Sydney Institute for Astronomy (SIfA), School of Physics, University of 
Sydney, NSW 2006, Australia}
\affiliation{Stellar Astrophysics Centre, Department of Physics and Astronomy, Aarhus 
University, Ny Munkegade 120, DK-8000 Aarhus C, Denmark}

\author{Shourya Khanna}
\affiliation{Sydney Institute for Astronomy (SIfA), School of Physics, University of 
Sydney, NSW 2006, Australia}

\begin{abstract}
The \kep\ mission has provided exquisite data to perform an ensemble asteroseismic analysis on evolved stars. In this 
work we systematically characterize solar-like oscillations and granulation for 16,094 oscillating red giants, using 
end-of-mission long-cadence data. We produced a homogeneous catalog of the frequency of maximum power (typical uncertainty 
$\sigma_{\nu_{\rm max}}$=1.6\%), the mean large frequency separation ($\sigma_{\Delta\nu}$=0.6\%), oscillation amplitude 
($\sigma_{\rm A}$=4.7\%), granulation power ($\sigma_{\rm gran}$=8.6\%), power excess width ($\sigma_{\rm width}$=8.8\%), 
seismically derived stellar mass ($\sigma_{\rm M}$=7.8\%), radius ($\sigma_{\rm R}$=2.9\%), and thus surface gravity 
($\sigma_{\log g}$=0.01 dex). Thanks to the large red giant sample, we confirm that red-giant-branch (RGB) and helium-core-burning 
(HeB) stars collectively differ in the distribution of oscillation amplitude, granulation power, and width of power excess, 
which is mainly due to the mass difference. The distribution of oscillation amplitudes shows an extremely sharp upper edge 
at fixed \numax, which might hold clues for understanding the excitation and damping mechanisms of the oscillation modes. We find that 
both oscillation amplitude and granulation power depend on metallicity, causing a spread of 15\% in oscillation amplitudes and 
a spread of 25\% in granulation power from [Fe/H]=-0.7 to 0.5 dex. Our asteroseismic stellar properties can be used as reliable 
distance indicators and age proxies for mapping and dating galactic stellar populations observed by \kep. They will also provide 
an excellent opportunity to test asteroseismology using Gaia parallaxes, and lift degeneracies in deriving atmospheric parameters 
in large spectroscopic surveys such as APOGEE and LAMOST.
\end{abstract}

\section{Introduction}
Red giants are bright, cool, and evolved stars that oscillate with amplitudes ranging from a few tens to thousands of 
parts per million and with characteristic oscillation timescales varying from hours up to months \citep{deridder09, huber11b, 
mosser12b, stello14}. Out of more than 196,000 stars observed by the \kep\ $Space\ Telescope$ \citep{borucki10,koch10b}, some 
19,000 oscillating red giants have so far been detected \citep{hekker11c, huber11b, stello13, huber14, mathur16, yu16}. The 
study of solar-like oscillations in giants has led to a number of breakthrough discoveries such as classification of the 
evolutionary stages of red giants \citep{bedding11, mosser12, stello13, mosser15, vrard16, elsworth17, hon17}, measurement 
of internal rotation \citep{beck12, deheuvels12, mosser12, deheuvels14} and possible detection of 
magnetic fields in radiative cores \citep{fuller15, stello16a, mosser17a}. It has also provided an excellent opportunity 
to implement Galactic archaeology \citep{miglio13, stello15, casagrande16, sharma16} and to characterize exoplanet properties 
\citep{huber13b, quinn15}.

Prior to the \kep\ mission, some analyses focusing on the seismic determination of stellar mass and radius were 
presented. For example, \citet{gilliland08} and \citet{stello09b} investigated the time series collected by the Hubble 
Space Telescope; \citet{stello08} worked with the star tracker of WIRE satellite; and \citet{kallinger10c} and 
\citet{mosser10} used data from the CoRoT telescope \citep{michel08}. Similar work has been done to derive the stellar 
properties for oscillating red giants observed by \kep, but only focusing on exoplanet host stars \citep{huber13} or using 
short datasets \citep{kallinger10, hekker11c}. This motivates us to study oscillations in red giants using the full four 
years of \kep\ data, aiming to provide a large and homogeneous catalog of seismic masses and radii. 

In order to determine stellar fundamental properties of red giants, three methods are widely used: the so-called direct method 
\citep{hekker11c}, grid-based modeling \citep{stello09b, kallinger10, huber13, chaplin14b}, and individual frequency modeling 
\citep{kallinger08a, dimauro11, deheuvels12, quinn15, dimauro16, li18}. Under the grid-based modeling method, 
atmospheric parameters and global seismic parameters are usually fitted to a grid of isochrones,  
which inevitably hold some model dependencies. This technique is efficient 
for main-sequence stars and subgiants but calls for additional efforts to have their evolutionary phases distinguished 
for red giants, since their evolutionary tracks converge in the Hertzsprung-Russell (H-R) diagram. The individual frequency 
analysis allows for the investigation of mass, age, and internal physical processes, such as overshooting and transport of  
angular momentum. But it rests on unambiguous identifications of a series of frequencies, which is only possible for 
high signal-to-noise ratio photometric time series. Furthermore, it is time-consuming to model individual frequencies for 
tens of thousands of red giants. The direct method makes use of seismic scaling relations, possibly 
with associated corrections, to efficiently determine stellar parameters. It has been tested theoretically and 
observationally, with a typical accuracy of $\sim$5\% and $\sim$10-15\% in radius and mass for red giants, respectively \citep{silva12, 
brogaard12, huber12, miglio12d, white13, gaulme16b, huber17}. We adopted the direct method to infer stellar fundamental properties 
in this work.

The primary goal of this work is to construct a large homogeneous catalog of global oscillation and granulation  
parameters and asteroseismic stellar masses, radii, and surface gravities for \kep\ oscillating red giants. We also 
attempt to investigate the mass and metallicity influence on oscillation amplitude, granulation power, and width of power excess. 
For this we use full-mission \kep\ data. Our sample consists of 16,094 \kep\ red-giant oscillators, representing the largest 
known sample so far to homogeneously perform an ensemble asteroseismic analysis.

\section{Sample Selection and Data Analysis}
\begin{figure*}
\begin{center}
\resizebox{!}{6cm}{\includegraphics{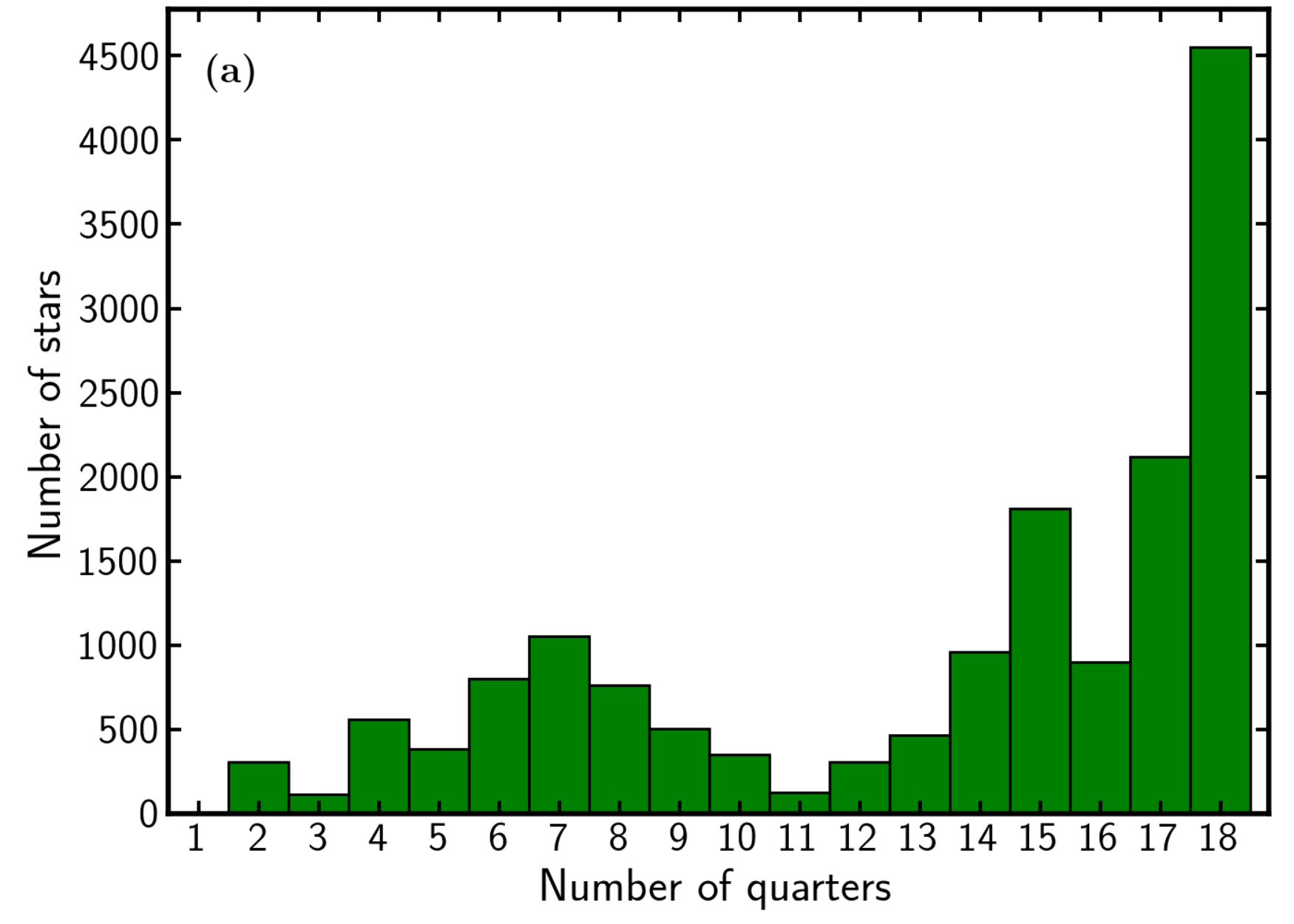}}
\resizebox{!}{6cm}{\includegraphics{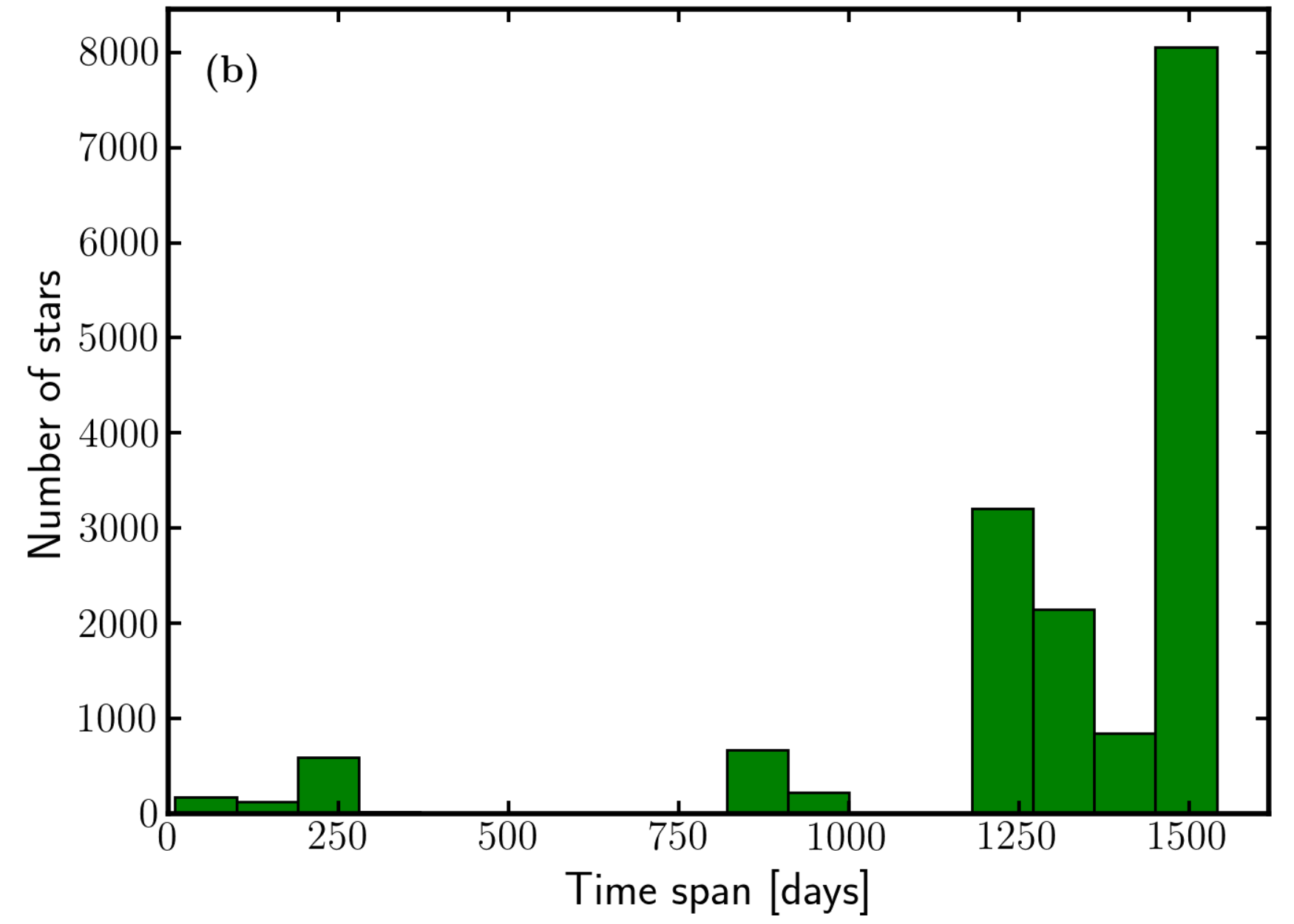}}\\
\resizebox{!}{6cm}{\includegraphics{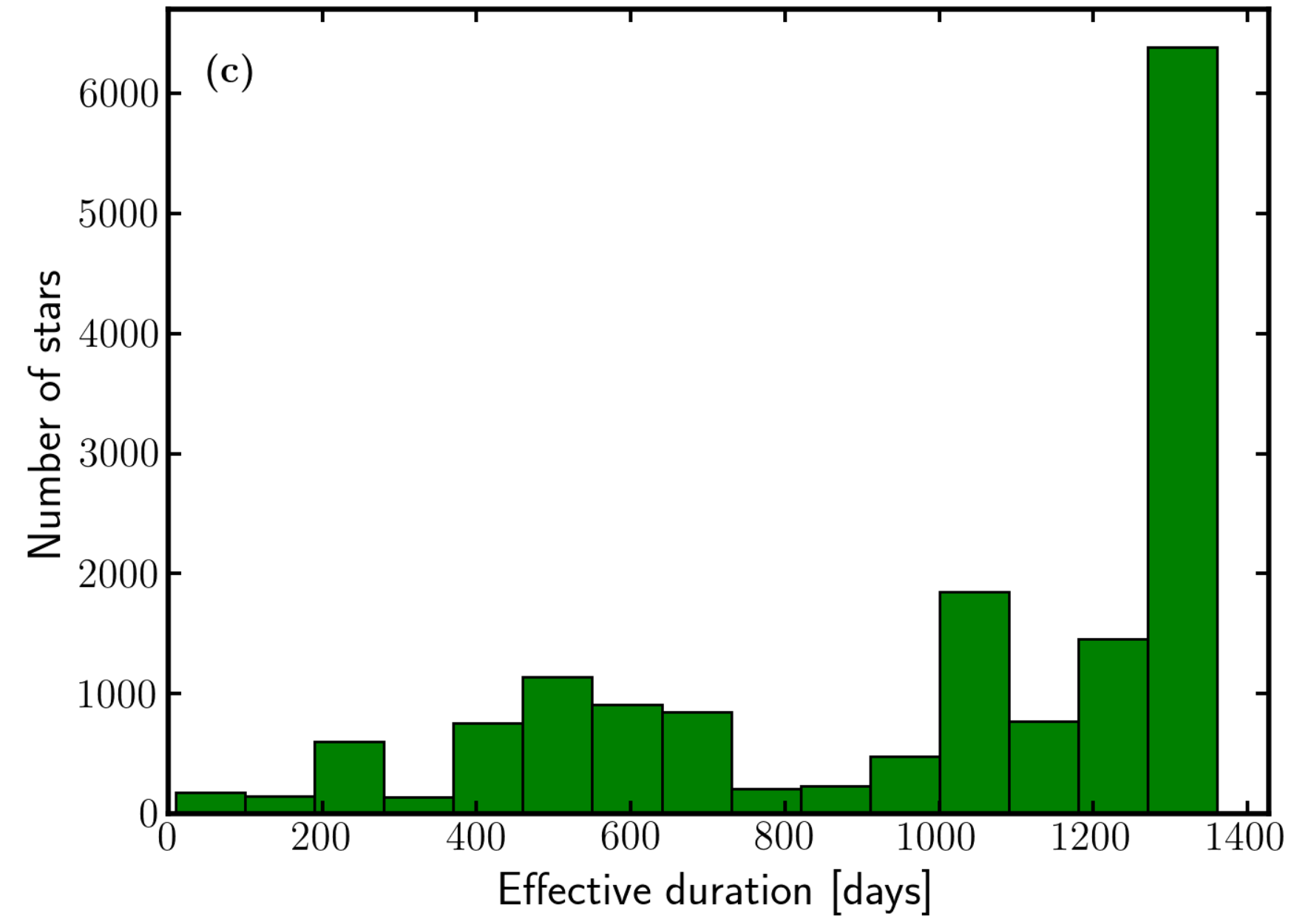}}
\resizebox{!}{6cm}{\includegraphics{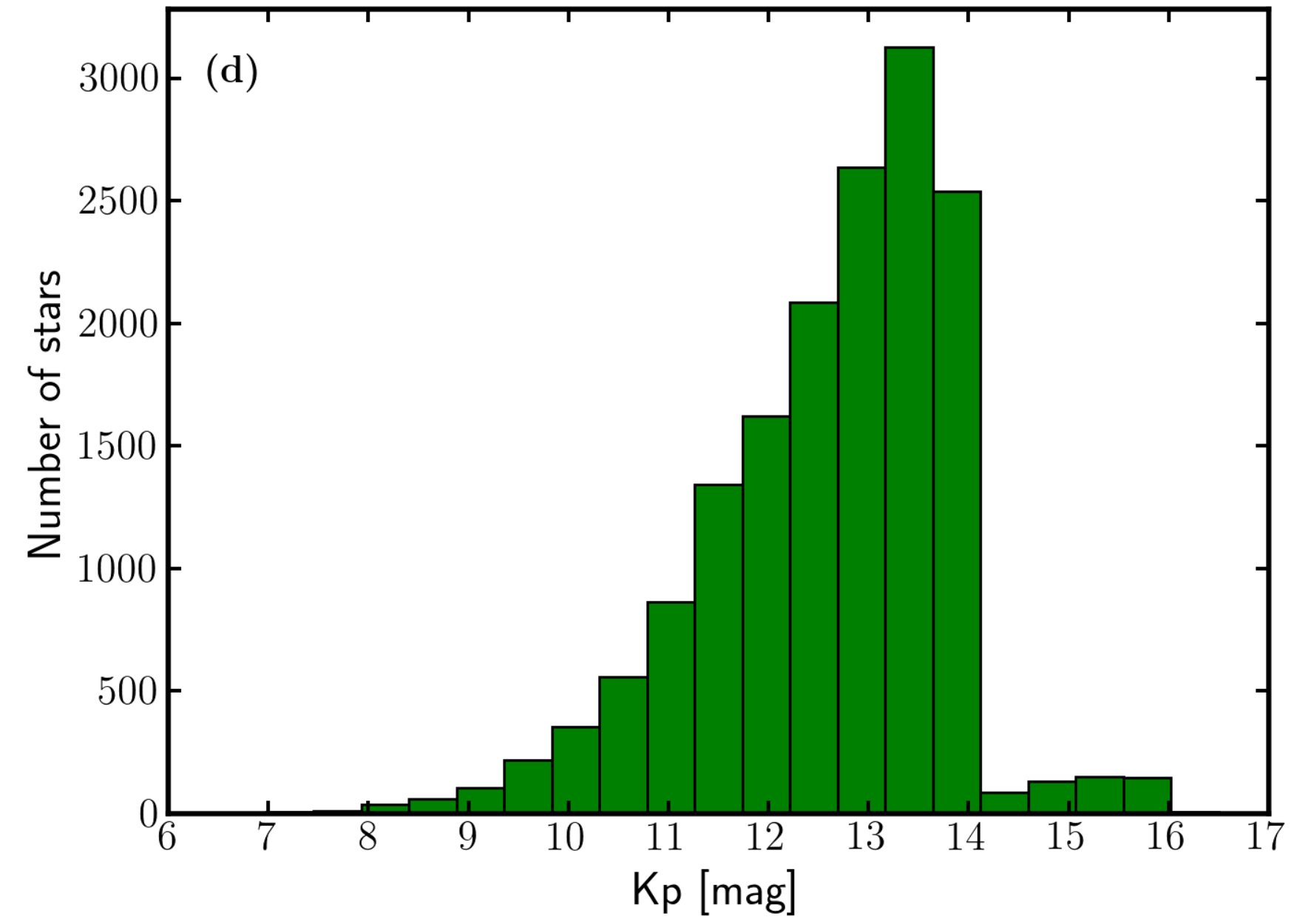}}\\
\caption{Distributions of time series length, including Q0 through Q17, and \kep\ magnitude for our entire sample of 16,094 oscillating 
red giants: \textbf{(a)} number of quarters; \textbf{(b)} time span; \textbf{(c)} effective duration (number of 
data points times the integration time of 29.4 minutes); and \textbf{(d)} \kep\ magnitude. The 
significant difference between panels \textbf{(a)} and \textbf{(b)} is due to some stars having big gaps in the time series.}
\label{figquarters}
\end{center}
\end{figure*}

\kep\ data are divided into quarters with a 10-day commissioning run, followed by the first 33-day quarter and 
subsequent 90-day regular quarters. In this work we make use of simple aperture photometry data collected in long-cadence 
mode. Figure \ref{figquarters} illustrates the histogram of duration of the long-cadence time series and the distribution 
of \kep\ magnitudes of our sample.

\begin{figure}
\begin{center}
\includegraphics[width=\columnwidth]{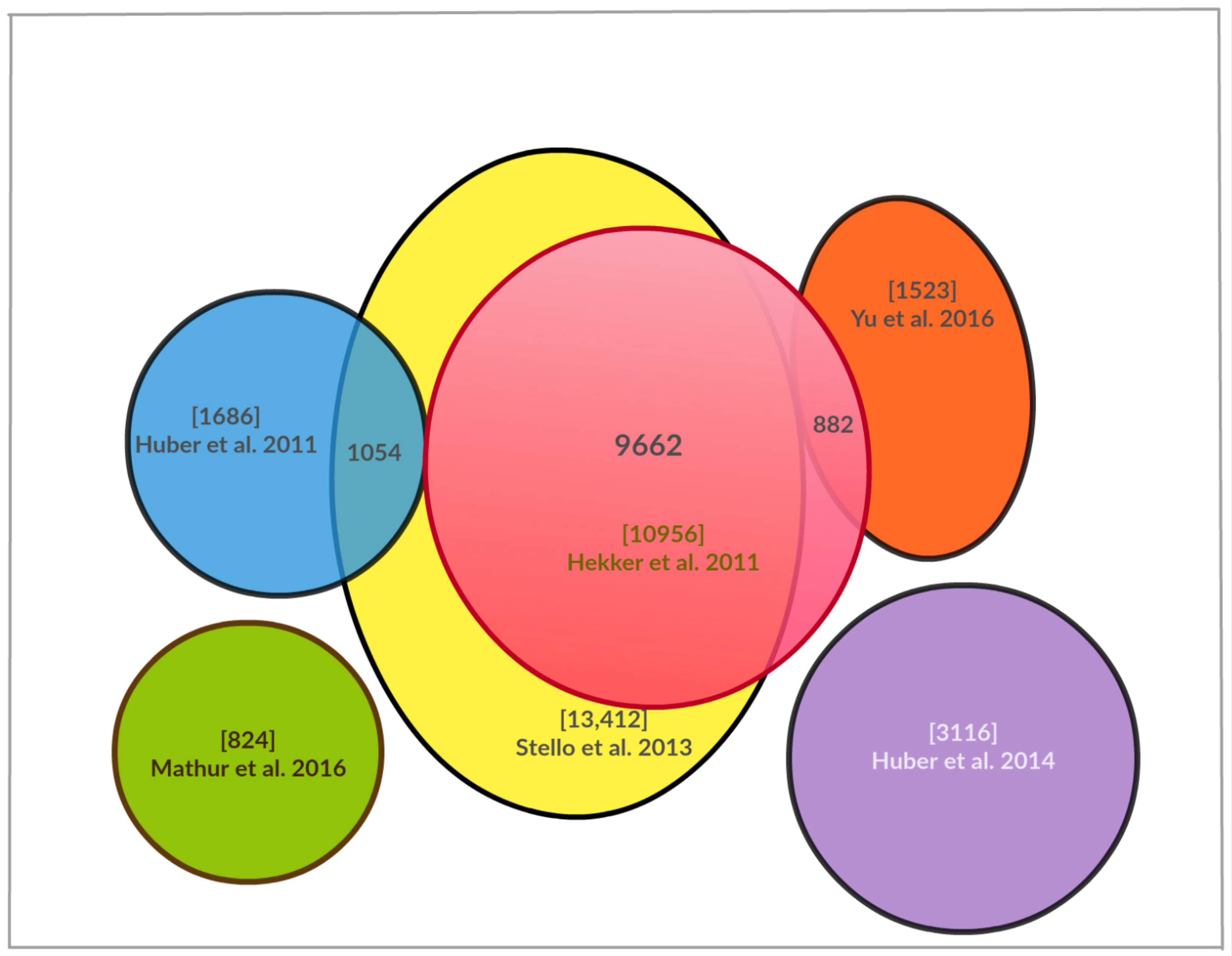}
\caption{Sample selection. The oscillating red giants constituting our entire sample are collected
        from six samples. The numbers in the square brackets show the original number 
        of stars in the corresponding source, while the unbracketed numbers indicate the number of common stars 
        in two overlapping samples. For clarity, overlaps of $\le50$ stars are not shown. 
        There are 16,094 oscillating red giants from these known samples constituting our entire sample, after applying a  
        cut $5\muHz<\numax<275\muHz$ and removing false and marginal detections.}
\label{figsample}
\end{center}
\end{figure}

Our sample is comprised of the known oscillating red giants from six published samples, as shown in the Venn diagram Figure \ref{figsample}. 
Some key properties of those six samples are summarized as follows:
\begin{itemize}
      \item \citet{hekker11c} conducted an asteroseismic characterization of over 16,000 
      red giants. For 10,956 red giants, oscillations were detected and stellar fundamental
      parameters were derived using the data recorded in the first quarter (Q1) of \kep\ data.  

      \item \citet{huber11b} used a sample of 1686 \kep\ targets consisting of dwarfs and giants  
      to test seismic scaling relations using long-cadence data spanning from Q0 to Q6 
      and short-cadence data from Q0 to Q4. 

      \item \citet{stello13} detected solar-like oscillations in 13,412 red giants, with the aim of  
      classifying evolutionary phase by measuring the period spacing of dipole modes identified with long-cadence 
      datasets from Q0 through Q8.

      \item \citet{huber14} presented a revised stellar properties catalog for 196,468 \kep\ targets, 
      and detected oscillations in 3114 stars that were unclassified in the \kep\ Input Catalog 
      \citep[KIC,][]{brown11}. For this sample, only the frequency of maximum oscillation power, 
      \numax, was measured.

      \item \citet{mathur16} discovered solar-like oscillations in over 800 faint and distant red 
      giants misclassified as dwarfs by the KIC.

      \item \citet{yu16} distinguished the real oscillation power excess from the aliased one in the power density spectrum and 
      discovered 626 new oscillating red giants that had been misclassified as subgiants in the KIC.
\end{itemize}

We excluded red giants with \numax\ $<$ 5 \muHz, resulting in a sample with \logg\ 
$\gtrsim$ 1.5 dex and luminosity log$(L/L\odot)$ $\lesssim$ 2.24 dex. Those excluded stars are expected to show oscillations 
in a few low radial-order acoustic modes \citep{stello14}. The seismic mass and radius inferred from scaling relations 
are likely to be biased for such star, as the scaling relations used are based on the asymptotic theory \citep{tassoul80, gough86}. 
We also removed dwarfs, subgiants, and stars with $\numax>275~\muHz$ due 
to the difficulty of fitting their power spectrum background. After visual inspection on our results, we removed outliers 
arising from wrong detections (non-oscillators) or marginal detections due to low signal-to-noise ratios. In the case where a star was 
analyzed in multiple literature samples, we adopted the one from the sample using the longest time series. 
Our final sample thus includes 133 stars from \citet{hekker11c}, 336 stars from \citet{huber11b}, 12,975 stars from \citet{stello13}, 
705 stars from \citet{huber14}, 606 stars from \citet{mathur16}, and 1339 stars from \citet{yu16}. This sample comprises 
16,094 oscillating red giants.

We corrected instrumental trends following the method described by \citet{garcia11}. Long-cadence light curves were stitched 
together, with safe-mode events removed and jumps corrected using a linear fit. A quadratic Savitzky-Golay high-pass filter was 
applied to remove instrument variability and low-frequency signals arising from stellar activity. We used an adaptive smoothing 
width, $d$, as a linear function of \numax\ (taken from the literature), following $d=0.61+0.04\numax$. Thus, the smoothing width 
varied from 0.8 to 13.5 \muHz\ when \numax\ increased from 5 \muHz\ to the Nyquist frequency.  A $4\sigma$-clipping was 
applied to remove outliers from the high-pass filtered light curves. 

The granulation and seismic parameters extracted in this work are the frequency of maximum oscillation power (\numax), the mean 
frequency separation of acoustic modes with the same angular degree and consecutive radial order (\Dnu), the oscillation amplitude 
per radial mode ($A$), the width of the power excess hump characterized by a Gaussian envelope, and 
granulation power measured at \numax. Specifically, the oscillation amplitude per radial mode is defined as \citep{kjeldsen08}: 
\begin{equation}
\label{amplitude}
A = \frac{\sqrt{\frac{H_{\rm env} \Delta\nu}{\it{c}}}}{\rm {sinc\left(\frac{\pi}{2} \frac{\nu_{max}}{\nu_{nyq}}\right)}} ,
\end{equation} 
where ${H_{\rm env}}$ is the height of the power excess 
hump, and $c$ is the effective number of modes per order, adopted as 3.04 \citep{bedding10, stello11}. We 
note that this value should in principle be adjusted for \mbox{dipole-mode} suppressed stars \citep{stello16a}, but this  
is beyond the scope of this paper. The attenuation of oscillation amplitude due to the integration of photons for every long-cadence 
interval (29.4 minutes) has been corrected with a sinc function \citep{huber10, murphy12, chaplin14}. Granulation power has also been 
corrected for this reason.

We used the SYD pipeline for extracting the granulation and seismic parameters mentioned above \citep{huber09}. The literature 
\numax\ values were assigned as initial guesses in the pipeline to model the power spectrum background. To obtain uncertainties 
for each parameter, we perturbed the power density spectrum 200 times with a $\chi^{2}$ distribution with two degrees of freedom, 
repeated the fitting procedures on each perturbed spectrum, and calculated the standard deviation of the resulting output parameter 
distributions as the formal uncertainties\citep{huber11b}. Results returned from the pipeline were verified by visual inspection 
for over 17,700 individual targets to remove a fraction (9.1\%) of wrong detections (non-oscillators) or marginal detections mainly 
due to low signal-to-noise ratios, resulting in a sample of 16,094 oscillating red giants.

\section{Determination of Seismic Stellar Fundamental Properties}
\begin{figure*}
\begin{center}
\includegraphics[width=\textwidth, keepaspectratio]{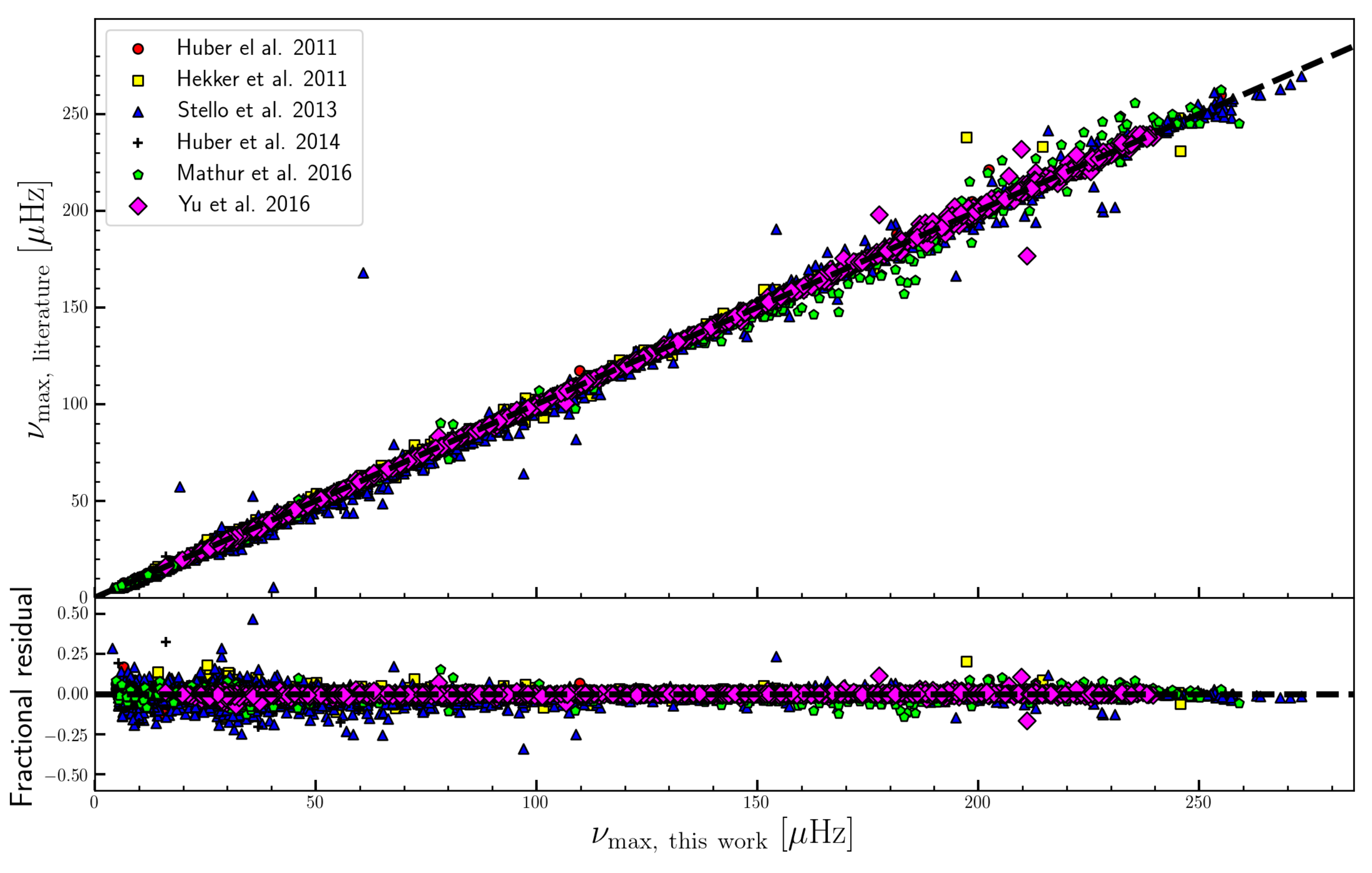}
\caption{\numax\ comparison between our measurements and those from the literature as shown in the legend. The black 
 dashed lines show the one-to-one relation in the top panel, and the fractional residual in the bottom panel in the sense 
 of $(\numax_{, \rm literature}-\numax_{, \rm this\ work})/\ \numax_{,\rm this\ work}$.}
\label{numaxcomp}
\end{center}
\end{figure*}

\begin{figure*}
\begin{center}
\includegraphics[width=\textwidth]{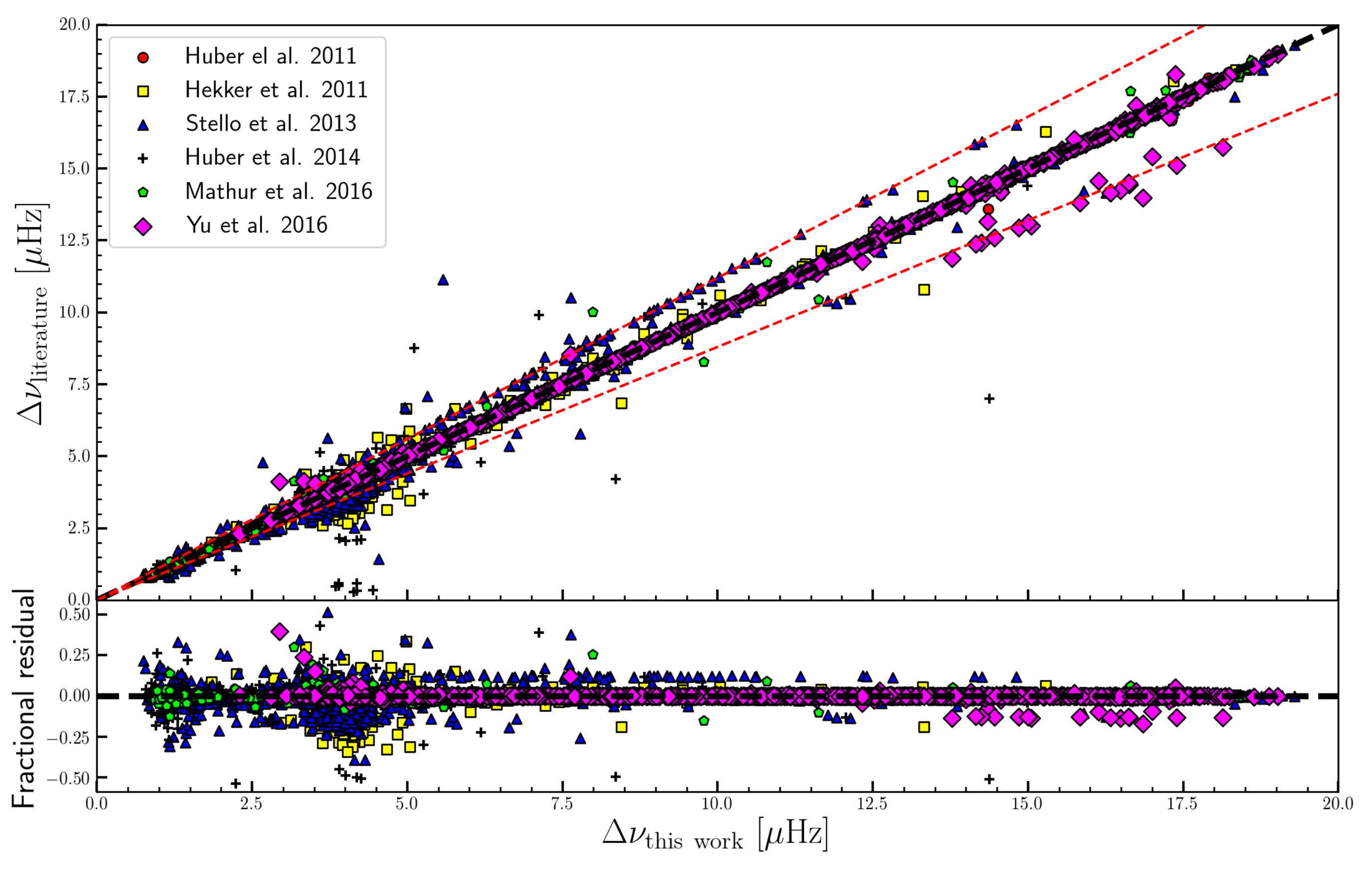}
\caption{Same as figure \ref{numaxcomp} but for \Dnu, except here we added two additional red dashed lines representing a 12\% 
deviation from the one-to-one relation, roughly equivalent to $\Dnu \pm \delta \nu _{02}$ 
(see the text).}
\label{dnucomp}
\end{center}
\end{figure*}

\subsection{Comparison of \numax\ Measurements}
Figure \ref{numaxcomp} shows a comparison of our \numax\ values to the six literature samples for 16,094 stars. 
Except for a few outliers, for which we have checked and confirmed our estimates by eye, our results show good agreement with the literature, 
displaying a median fractional residual of 0.2\% and a scatter of 3.5\%. We note that the \numax\ estimates are not measured with totally 
independent methods. \citet{huber11b}, \citet{stello13}, \citet{huber14} and \citet{yu16} all used versions of the SYD pipeline 
(the main difference in measuring \Dnu, see the details in Sec. \ref{dnuvaluescomp}), which are different from those employed 
by \citet{hekker11c} and \citet{mathur16}. The comparison of \numax\ estimates from our work with those from \citet{mathur16} 
for 606 common stars shows a median fractional residual of $-0.7$\% and a scatter of 4.5\%. The comparison 
with the \citet{hekker11c} for 10,727 common starsshows a median fractional residual of $-0.8$\% and a 
scatter of 10.7\%. The larger scatter between our results and \citet{hekker11c} is mainly due to the fact that \citet{hekker11c} used 
only Q1 time series, resulting in some incorrectly measured \numax\ values, presumably due to low signal-to-noise ratio spectra. 

From the bottom panel we note that a significant spread occurs in the \numax\ range, corresponding to red clump stars, roughly 
around 30 \muHz, and to high-luminosity red giants less than approximately 10 \muHz. We found that the fractional residuals rise with 
decreasing \numax. The significant spread at \numax\ $\approx$ 30 \muHz\ is an apparent effect due to a larger number of stars 
in this parameter range, which is also present in the \numax\ comparison performed by \citet{hekker11}. The significant spread 
among stars with \numax\ less than 10 \muHz\ is associated with larger measurement uncertainties due to the higher frequency 
resolution required for these stars. Those high-luminosity red giants were generally observed with 25\% fewer data points than 
stars with \numax\ $\sim$200 \muHz, as a result of \kep\ target selection effects \citep{batalha10}. The second 
reason is related to the fewer detectable modes for measuring \numax. These stars have only approximately three orders of detectable modes 
\citep{stello14}, which, due to the stochastic nature of the mode excitation, can result in large uncertainties in \numax\ measurements. 
Another spread arises from the \citet{mathur16} sample, as shown with the green filled pentagons. This presumably arises from the fact 
that those stars are distant low-luminosity red giants. The use of different methods to measure global oscillation parameters also 
contributes to the scatter. For further details on the comparison of different methods for the determination of 
global seismic parameters, we refer the reader to \citet{hekker11, verner11}, and \citet{hekker12}.

\subsection{Comparison of \Dnu\ Measurements}
\label{dnuvaluescomp}
Figure \ref{dnucomp} shows the comparison of \Dnu\ measurements from different methods for 16,094 stars. 
The fractional residuals show two groups of stars, with 
the first one lying around \Dnu\ $\simeq$ 1.3 \muHz\ and the second one between approximately 3.3 and 5~\muHz. The first 
group associated with high-luminosity red giants is related to low frequency resolution and to having few orders of detectable oscillation 
frequencies. The second group mainly corresponds to red clump stars, which have less regular and less clean power spectra. A combination  
of broad linewidths, mixed modes, rotational splittings, and acoustic glitches can lead to large autocorrelations in a broad frequency 
range around the real \Dnu\ values. Those effects were also present in the comparison work presented by \citet{hekker11}. Overall, our results 
are consistent with the literature, with a median fraction residual of 0.01\% and a scatter of 4.2\%. The comparison of our \Dnu\ estimates 
with the different method of \citet{mathur16} or 606 common stars shows an absolute median fractional residual 
of $<$0.01\% and a scatter of 3.8\%. The comparison with \citet{hekker11c} for 10,727 common stars shows a median fraction 
residual of 0.8\% and a scatter of 14.4\%.

There are a number of stars (blue triangles) from the \citet{stello13} sample with measured \Dnu\ values systematically 
12\% larger than our measurements (red dashed line above the one-to-one relation). Those \Dnu\ values correspond to the frequency differences 
$\delta \nu = \nu_{n+1, l=0}-\nu_{n-1, l=2} = \Dnu+\delta \nu_{02}$. This is confirmed by comparing the measurements of the 
small frequency separation $\delta \nu _{02}$, which roughly has a fixed ratio $\delta \nu _{02}/ \Dnu\ = 0.121$, as measured by 
\citet{huber10}. There are also some stars from the \citet{stello13} and \citet{yu16} samples, with \Dnu\ values systematically 12\% 
smaller than the ones determined in this work (red dashed line below the one-to-one relation). Those values are the frequency 
differences $\delta \nu = \nu _{n, l=2}-\nu _{n, l=0} = \Dnu-\delta \nu _{02}$. These incorrect \Dnu\ measurements from the literature 
also contribute to the scatter seen for the two groups of stars.

To understand the errors in \Dnu\ measurements in \citet{stello13} and \citet{yu16}, we note that, 
in the original SYD pipeline, the resulting \Dnu\ value was determined to be whichever of the 10 highest peaks of the autocorrelation 
function is closest to the predicted \Dnu. This suggests that a less accurately \textit{predicted} \Dnu\, , based 
on the \numax-\Dnu\ power-law relation, could lead to an incorrectly 
\textit{measured} \Dnu\ value, such as, $\Dnu - \delta \nu _{02}$ or $\Dnu + \delta \nu _{02}$. Given that the real \Dnu\ value generally 
corresponds to a larger autocorrelation compared to $\Dnu - \delta \nu _{02}$ and $\Dnu + \delta \nu _{02}$, we weighted the 
autocorrelation functions using a Gaussian function centered at the predicted \Dnu\ with a width of 30\% or 70\% of the predicted 
\Dnu. The higher width (70\%) was specifically assigned to stars with \numax\ in the range $15\ \muHz <\numax<60\ \muHz$, which is 
mostly occupied by red clump stars. The lower width (30\%) was applied to stars with \numax\ outside this range, and its reliability 
was confirmed by noting that \Dnu\ values of RGB and secondary clump stars are much less sensitive to the selected width. The highest weighted 
peak was adopted as the resulting \Dnu\ value. We found this method works well to correct those 12\%-biased measurements. 
All measured global oscillation parameters were visually verified and are listed in Table \ref{seisparatable}.

\subsection{Asteroseismic Scaling Relations}
\label{bg}
\citet{brown91} suggested that \numax\ would scale with the acoustic cutoff frequency and hence be related to 
stellar fundamental properties, as given by \citet{KB95}, as follows:
\begin{equation}  
\label{numaxscale} 
\frac {\numax}{\numax _{\sun}} \simeq \left(\frac{M}{\rm M_{\sun}}\right)
\left(\frac{R}{\rm R_{\sun}}\right)^{-2}\left(\frac{T_{\rm eff}}{\rm T_{\rm eff, \sun}}\right)^{-1/2}.
\end{equation}
Here, $\numax_{,\sun}=3090 \pm 30\ \muHz$, $\rm T_{\rm eff,\sun} = 5777 \rm K$.
The other widely used scaling relation is related to the large frequency separation, \Dnu, which probes the sound 
speed profile and is proportional to the square root of the mean stellar density, as proposed by \citet{ulrich86}, as
follows:
\begin{equation}
\label{dnuscale}
\frac {\Delta \nu}{\Delta \nu _{\sun}} \simeq \left(\frac {M}{\rm M_{\sun}}\right)^{1/2} 
\left(\frac {R}{\rm R _{\sun}}\right)^{-3/2},
\end{equation} 
where $\Delta \nu _{\sun} = 135.1 \pm 0.1 \muHz$. The solar seismic reference values are obtained 
by analyzing the data collected by $SOHO$/VIRGO \citep{frohlich97} in the same way as the analyzed \kep\ data \citep{huber11}.  
By rearranging the scaling relations, stellar mass, $M$, radius, $R$, and surface gravity, $g$, can be readily derived as follows:
\begin{equation}
\frac{M}{\rm M_{\sun}} \simeq \left(\frac {\numax}{f_{\numax} \numax, _{\sun}} \right)^{3}
\left(\frac{\Delta\nu}{f_{\Delta\nu} \Dnu _{\sun}}\right)^{-4}
\left(\frac{T_{\rm eff}}{\rm T_{\rm eff,\sun}}\right)^{3/2},
\label{massscaling}
\end{equation}
\begin{equation}
\frac{R}{\rm R_{\sun}} \simeq \left(\frac {\numax}{f_{\numax} \numax, _{\sun}} \right)
\left(\frac{\Dnu}{f_{\Delta\nu} \Dnu _{\sun}}\right)^{-2}
\left(\frac{\teff}{\rm T_{eff,\sun}}\right)^{1/2},
\label{radiusscaling}
\end{equation}
\begin{equation}
\label{loggseiseq}
\frac {g}{\rm{g}_{\sun}} \simeq \frac {\numax}{f_{\numax} \numax, _{\sun}}\left(\frac{\teff}{\rm T_{\rm eff,\sun}}\right)^{1/2}.
\end{equation}
Here, $f_{\numax}$ and $f_{\Delta\nu}$ are the potential correction factors for the \numax\ and \Dnu\ scaling relations, respectively.

\begin{table*}[t]
\begin{footnotesize}
\begin{centering}
\caption{Stellar Global Oscillation Parameters}
\resizebox{\textwidth}{!}{\begin{tabular}{rllllllll}
\hline
\hline
KIC & Kp   & Length      & Length & \numax      & \Dnu       & $A$   & Width   & Gran \\
    & (mag)&  (quarters) & (days) &   (\muHz)   &  (\muHz)   & (ppm) & (\muHz) & ($\rm{ppm^2}$/\muHz) \\
\hline
   2570518  &     14.72  &   17  &    1308.6  &     46.12\ (0.75)  &     4.934\ (0.012)  &     98.5\ ( 5.4)  &     16.7\ ( 1.3)  &     1312.6\ (335.8)\\
   4682420  &     13.12  &   16  &    1221.3  &    128.72\ (1.27)  &    10.350\ (0.014)  &     32.8\ ( 1.8)  &     44.7\ (10.2)  &       63.8\ ( 10.6)\\
   4946632  &     13.43  &   18  &    1318.2  &    199.32\ (1.08)  &    15.082\ (0.074)  &     45.8\ ( 3.5)  &     47.8\ ( 4.5)  &       25.8\ ( 12.5)\\
   5340720  &     12.84  &   14  &    1023.4  &     94.88\ (0.71)  &     8.901\ (0.019)  &     57.7\ ( 2.9)  &     26.6\ ( 1.6)  &      311.9\ ( 76.8)\\
   5446355  &     12.79  &   18  &    1317.8  &      8.04\ (0.25)  &     1.266\ (0.029)  &    277.3\ (22.2)  &      3.4\ ( 0.5)  &    92604.4\ (6767.2)\\
   6197448  &     11.83  &   15  &    1138.8  &     50.45\ (1.18)  &     4.927\ (0.018)  &     58.1\ ( 2.8)  &     21.4\ ( 1.7)  &     1093.5\ ( 61.1)\\
   6429836  &     13.83  &    5  &     371.3  &     38.82\ (0.53)  &     4.298\ (0.025)  &    120.1\ ( 6.4)  &     13.8\ ( 1.9)  &     2132.0\ (283.5)\\
   6435899  &     13.53  &    6  &     380.7  &     21.50\ (0.42)  &     3.009\ (0.104)  &    180.5\ ( 9.0)  &      7.6\ ( 1.0)  &     9218.7\ (1309.2)\\
   6756156  &     13.75  &   13  &    1013.9  &    160.46\ (1.10)  &    13.673\ (0.023)  &     43.7\ ( 2.3)  &     43.4\ ( 2.7)  &       75.5\ ( 22.0)\\
   7445517  &     12.93  &   17  &    1308.5  &     64.39\ (0.80)  &     5.905\ (0.014)  &     58.9\ ( 2.9)  &     21.5\ ( 1.4)  &      654.9\ ( 84.6)\\
   8265154  &     13.76  &   17  &    1308.9  &    208.08\ (1.79)  &    16.589\ (0.050)  &     $\textendash$ &   $\textendash$ &     $\textendash$\\
   8509198  &     13.80  &   10  &     777.2  &    108.95\ (1.29)  &     9.025\ (0.017)  &     33.3\ ( 2.1)  &     35.3\ ( 3.1)  &      145.8\ ( 32.4)\\
   9285761  &     12.41  &   16  &    1221.0  &     53.23\ (0.47)  &     5.478\ (0.017)  &     84.7\ ( 4.5)  &     17.5\ ( 1.1)  &     1115.6\ ( 48.7)\\
   9475300  &     12.75  &   18  &    1318.0  &     64.61\ (0.48)  &     6.303\ (0.011)  &     73.6\ ( 2.8)  &     20.6\ ( 1.1)  &      791.8\ ( 48.0)\\
  10318430  &     12.03  &   18  &    1318.0  &    154.84\ (0.91)  &    12.985\ (0.032)  &     39.0\ ( 1.3)  &     45.9\ ( 2.3)  &       46.3\ (  8.0)\\
  10420502  &     12.78  &    7  &     467.3  &     32.28\ (0.65)  &     4.028\ (0.052)  &    135.2\ ( 5.7)  &     13.6\ ( 1.9)  &     4911.1\ (444.2)\\
  10675935  &     12.95  &   15  &    1055.4  &     49.55\ (1.20)  &     4.897\ (0.069)  &     61.1\ ( 2.4)  &     21.6\ ( 1.8)  &     1008.4\ ( 49.9)\\
  11026843  &     11.10  &   17  &    1235.3  &     30.27\ (0.63)  &     3.868\ (0.021)  &    132.1\ ( 5.4)  &     12.4\ ( 1.4)  &     5139.3\ (322.1)\\
  11600442  &      8.85  &   15  &    1044.7  &     69.64\ (0.87)  &     6.048\ (0.044)  &     43.4\ ( 1.7)  &     24.8\ ( 1.5)  &      439.2\ ( 35.3)\\
  12555883  &     12.57  &   14  &    1052.8  &     54.36\ (0.82)  &     5.242\ (0.014)  &     71.4\ ( 3.1)  &     21.0\ ( 1.6)  &     1200.1\ ( 75.5)\\
\hline
\end{tabular}}
\label{seisparatable}
\flushleft Note. The length of the dataset, in numbers of quarters (third column) and in days (fourth column), includes 
the 10-day commissioning run (Q0). The oscillation amplitude per radial mode, power excess width, and granulation power at \numax\   
can be found in the last three columns, for the stars with $\numax\leq200\ \muHz$. The values 
in the brackets represent absolute formal uncertainties. (This table is available in its entirety in a 
machine-readable form in the online journal. A portion is shown here for guidance regarding its form and content.)
\end{centering}
\flushleft
\end{footnotesize}
\end{table*}

The stellar radius inferred from scaling relations has been tested to hold within $\sim$5\% for both dwarfs 
and giants using parallaxes, eclipsing binaries, cluster stars, and optical interferometry \citep{silva12, brogaard12, 
huber12, white13, huber17}. The stellar mass from the direct method has been tested to have $\sim$10-15\% uncertainties \citep{miglio12d, 
gaulme16b}.  

Some efforts have been made to mitigate the possible systematics of the scaling relations. \citet{Miglio12b} proposed that a 2.7
\%(1.9\%) correction factor of \Dnu\ for red giants in NGC 6791 (NGC 6819) is necessary to minimize the difference in radius 
evaluated from the scaling relation and independent measurements of luminosity and effective temperature. 
\citet{mosser13} pointed out that the observed large frequency separation, $\Delta\nu_{\rm obs}$, measured at radial orders that were not high 
enough, is not equivalent to its asymptotic approximation, $\Delta\nu_{\rm as}$, linked to the mean density of the star. A 
second-order term, for describing curvatures in the \'echelle diagram used for characterizing solar-like oscillations, was accounted 
for to revise the \Dnu\ scaling relation. \citet{hekker13b}, however, stated that the correction to the scaling relations is overestimated, by  
comparing $\Delta\nu_{\rm obs}$ and $\Delta\nu_{\rm as}$ from stellar models. \citet{yildiz16} argued that the \Dnu\ scaling relation also depends 
on the adiabatic exponent at the surface, $\Gamma_{1s}$, but the application of their correction method is restricted to main-sequence stars.
\citet{white11, guggenberger16}, and \citet{guggenberger17} proposed corrections to the \Dnu\ scaling relations based on stellar models, 
including a dependence on temperature, metallicity, and mass. However, these corrections do not include HeB red giants, which 
make up roughly half of our sample. A similar approach suggested by \citet{sharma16} is to use a \Dnu\ correction factor, $f_{\Delta\nu}$, 
which is a function of metallicity, \teff, \logg, and evolutionary phase. The correction factor is obtained by interpolation in grids of models 
for $-3<\rm{[Fe/H]}<0.4$ and $0.8<\rm{M/\msun}<4.0$. We used this method to calibrate the \Dnu\ scaling relation. The \numax\ 
calibration is more difficult since it cannot be calculated theoretically so far \citep{belkacem11}. Thus, we set 
$f_{\nu_{\rm max}}=1.0$ in this work.

\begin{table*}[t]
\begin{center}
\caption{Stellar Fundamental Properties}  
\resizebox{\textwidth}{!}{\begin{tabular}{rllllllllll}
\hline
\hline
&&&&\multicolumn{2}{c}{\bf No \Dnu\ correction}&\multicolumn{2}{c}{\bf \Dnu\ corrected, RGB}&\multicolumn{2}{c}{\bf \Dnu\ corrected, Clump}  \\ 
\cline{5-6} \cline{7-8}  \cline{9-10}  
KIC & $\teff$ & $\logg$  & $\feh$ & $M$     & $R$     & $M$       & $R$     & $M$     & $R$    & Phase\\
    &  (K)    & (c.g.s.) &        & (\msun) & (\rsun) & (\msun)   & (\rsun) & (\msun) & (\rsun)\\
\hline
   2570518  &  4531\ ( 80) & 2.559\ (0.009) &  0.360\ (0.150) & 1.30\ (0.09) &  9.91\ (0.24) & 1.17\ (0.08) &  9.41\ (0.22) & 1.30\ (0.09) &  9.90\ (0.24) & 1 \\
   4682420  &  4827\ ( 80) & 3.019\ (0.007) &  0.230\ (0.150) & 1.60\ (0.09) &  6.49\ (0.13) & 1.53\ (0.08) &  6.33\ (0.12) & 1.63\ (0.09) &  6.54\ (0.13) & 1 \\
   4946632  &  4773\ ( 80) & 3.206\ (0.006) &  0.390\ (0.150) & 1.30\ (0.07) &  4.70\ (0.09) & 1.25\ (0.06) &  4.61\ (0.09) & 1.30\ (0.07) &  4.70\ (0.09) & 1 \\
   5340720  &  4995\ (146) & 2.894\ (0.008) & -0.298\ (0.300) & 1.24\ (0.08) &  6.58\ (0.15) & 1.18\ (0.07) &  6.43\ (0.14) & 1.25\ (0.08) &  6.63\ (0.15) & 1 \\
   5446355  &  4336\ ( 80) & 1.791\ (0.015) & -0.070\ (0.150) & 1.49\ (0.21) & 25.68\ (1.48) & 1.31\ (0.18) & 24.10\ (1.36) & 1.46\ (0.20) & 25.41\ (1.46) & 1 \\
   6197448  &  4756\ ( 80) & 2.609\ (0.012) &  0.330\ (0.150) & 1.84\ (0.15) & 11.14\ (0.33) & 1.74\ (0.15) & 10.83\ (0.32) & 1.87\ (0.16) & 11.22\ (0.33) & 2 \\
   6429836  &  4758\ (141) & 2.495\ (0.010) &  0.065\ (0.300) & 1.45\ (0.11) & 11.27\ (0.31) & 1.34\ (0.10) & 10.84\ (0.29) & 1.47\ (0.11) & 11.34\ (0.31) & 1 \\
   6435899  &  4832\ (100) & 2.242\ (0.011) & -0.410\ (0.300) & 1.05\ (0.17) & 12.83\ (0.95) & 0.94\ (0.14) & 12.14\ (0.88) & 1.04\ (0.16) & 12.80\ (0.95) & 2 \\
   6756156  &  5070\ (151) & 3.125\ (0.008) & -0.526\ (0.300) & 1.10\ (0.07) &  4.75\ (0.10) & 1.06\ (0.06) &  4.67\ (0.10) & 1.12\ (0.07) &  4.80\ (0.11) & 1 \\
   7445517  &  4756\ ( 80) & 2.715\ (0.008) &  0.170\ (0.150) & 1.85\ (0.11) &  9.90\ (0.21) & 1.73\ (0.10) &  9.56\ (0.20) & 1.89\ (0.11) & 10.00\ (0.21) & 1 \\
   8265154  &  4967\ (149) & 3.234\ (0.009) & -0.414\ (0.300) & 1.07\ (0.07) &  4.14\ (0.10) & 1.01\ (0.07) &  4.03\ (0.09) & 1.09\ (0.07) &  4.18\ (0.10) & 1 \\
   8509198  &  4993\ (162) & 2.954\ (0.010) & -0.122\ (0.300) & 1.77\ (0.12) &  7.35\ (0.18) & 1.69\ (0.12) &  7.19\ (0.18) & 1.81\ (0.13) &  7.43\ (0.18) & 1 \\
   9285761  &  4803\ ( 80) & 2.634\ (0.007) & -0.170\ (0.150) & 1.43\ (0.08) &  9.55\ (0.19) & 1.30\ (0.07) &  9.11\ (0.18) & 1.45\ (0.08) &  9.60\ (0.19) & 1 \\
   9475300  &  4783\ ( 80) & 2.717\ (0.007) & -0.080\ (0.150) & 1.45\ (0.07) &  8.74\ (0.16) & 1.33\ (0.07) &  8.36\ (0.15) & 1.47\ (0.07) &  8.79\ (0.16) & 1 \\
  10318430  &  5329\ (151) & 3.121\ (0.008) & -0.307\ (0.300) & 1.31\ (0.08) &  5.21\ (0.11) & 1.36\ (0.08) &  5.32\ (0.11) & 1.33\ (0.08) &  5.26\ (0.11) & 1 \\
  10420502  &  4735\ ( 80) & 2.414\ (0.010) &  0.030\ (0.150) & 1.07\ (0.10) & 10.64\ (0.39) & 1.00\ (0.09) & 10.27\ (0.37) & 1.08\ (0.10) & 10.67\ (0.39) & 2 \\
  10675935  &  5129\ (154) & 2.617\ (0.013) & -0.323\ (0.300) & 2.00\ (0.22) & 11.50\ (0.49) & 1.95\ (0.21) & 11.37\ (0.48) & 2.07\ (0.22) & 11.70\ (0.50) & 2 \\
  11026843  &  5009\ (150) & 2.398\ (0.012) &  0.067\ (0.300) & 1.13\ (0.10) & 11.13\ (0.35) & 1.15\ (0.10) & 11.23\ (0.35) & 1.14\ (0.10) & 11.16\ (0.35) & 2 \\
  11600442  &  5205\ (158) & 2.768\ (0.010) &  0.324\ (0.300) & 2.44\ (0.18) & 10.68\ (0.30) & 2.63\ (0.20) & 11.09\ (0.32) & 2.49\ (0.19) & 10.79\ (0.31) & 2 \\
  12555883  &  4648\ (137) & 2.636\ (0.010) &  0.533\ (0.300) & 1.73\ (0.13) & 10.48\ (0.27) & 1.61\ (0.12) & 10.11\ (0.26) & 1.77\ (0.13) & 10.60\ (0.28) & 1 \\
\hline
\end{tabular}}
\label{stellarparatable}
\flushleft Note. \teff\ and \feh\ are collected from \citet{mathur17}, while surface gravity is seismically derived 
from this work. Three solutions of stellar mass and radius are provided, corresponding to those with and without \Dnu\ 
corrections. Evolutionary phases are also given in the last column, 2 for HeB, 1 for RGB, and 0 for unclassified phase, 
with which stellar mass and radius can be obtained readily. For example, KIC 2570518 is an RGB star, whose mass and 
radius are respectively 1.17$\pm$0.08\msun\ and 9.41$\pm$0.22\rsun\ after \Dnu\ correction. The values 
in the brackets represent absolute uncertainties. (This table is available in its entirety in a machine-readable form 
in the online journal. A portion is shown here for guidance regarding its form and content.)
\end{center}
\flushleft
\end{table*}

\subsection{Determination of Stellar Parameters}
\label{params}
\begin{figure*}
\begin{center}
\includegraphics[width=\textwidth, height=14cm, keepaspectratio]{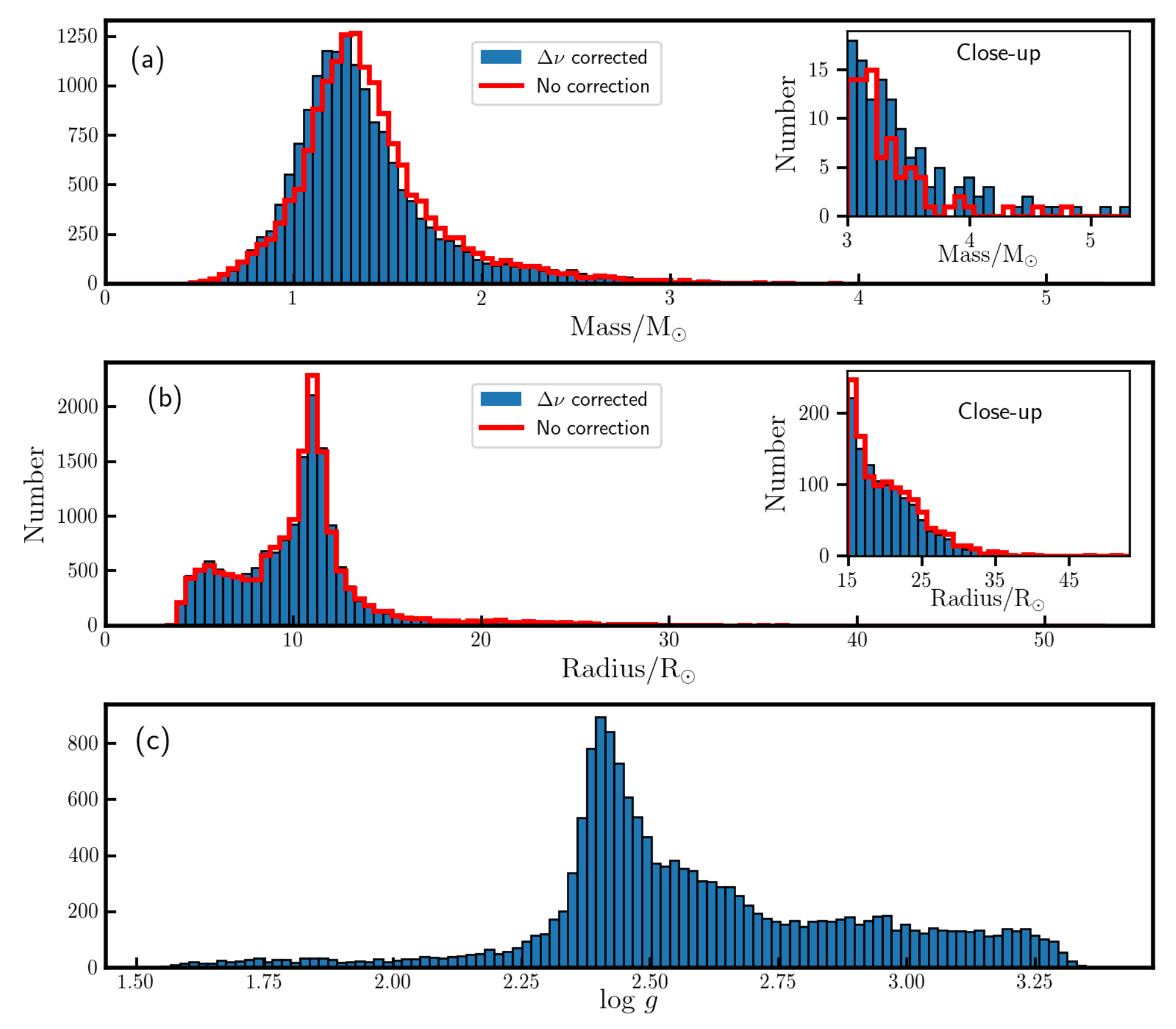}
\caption{Histograms of seismically inferred mass, radius, and surface gravity. The distributions of \textbf{(a)} mass 
and \textbf{(b)} radius in blue accouter for the \Dnu\ correction using the scheme proposed by \citet{sharma16}. 
For comparison, we also show the mass and radius distributions without the \Dnu\ correction in red. The surface gravity distribution is 
plotted in panel (\textbf{c}). The insets show a close-up of the distributions of high-mass and high-radius stars.}
\label{figmassradius}
\end{center}
\end{figure*}

We combine \numax\ and \Dnu\ from this work with effective temperatures from \citet{mathur17} to compile a 
homogeneous catalog of seismically derived stellar mass, radius, and therefore surface gravity. We adopted 
the model-based method proposed by \citet{sharma16} to correct \Dnu\ and applied the direct method for deriving mass, radius, and \logg. Since 
the correction factor $f_{\Delta\nu}$ is different for RGB and HeB stars, the classification of evolutionary stage is required. 
For this we used the results from \citet{bedding11}, \citet{stello13}, \citet{mosser14}, \citet{vrard16}, \citet{elsworth17}, 
and \citet{hon17}. 

\begin{figure*}
\begin{center}
\includegraphics[width=\textwidth, height=14cm, keepaspectratio]{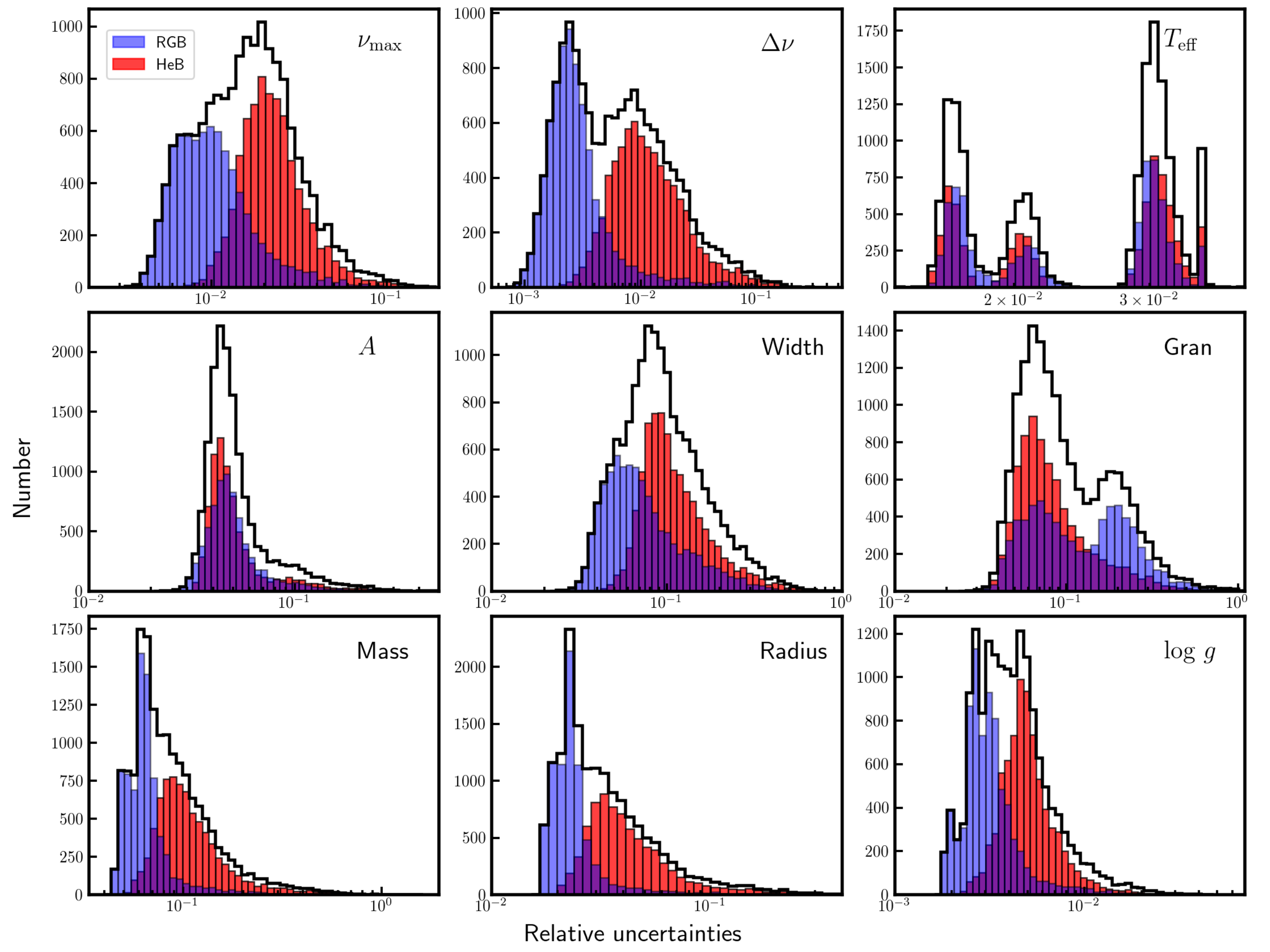}
\caption{Relative uncertainty distributions of global oscillation parameters and stellar fundamental properties for 16,094 stars. When deriving the 
uncertainties of mass, radius, and \logg, we have added 0.5\% and 1\% uncertainties in quadrature to the formal uncertainties of \Dnu\ 
and \numax, respectively. The blue and red bars show overlapping distributions of RGB and HeB stars, respectively, while the black histogram displays 
the sum of all the stars, including the 706 stars with an unclassified (RGB or HeB) evolutionary phase.}
\label{uncertainties}
\end{center}
\end{figure*}

We provide three solutions of mass and radius estimates in Table \ref{stellarparatable}, one with \Dnu\ corrected 
assuming all the targets are RGB stars, one with \Dnu\ corrected but assuming all the targets are HeB stars, and the 
third one without any \Dnu\ correction. We recommend using mass and radius estimates with the \Dnu\ 
correction taken into account.  Mass and radius values can be readily obtained from Table \ref{stellarparatable} 
if the evolutionary stage is known. The recommended evolution phases are given in the last column  
of Table \ref{stellarparatable}. Given the length of the time series and the oscillation signal-to-noise ratio for the 
targets in those samples, we gave the highest reliability to \citet{hon17} (which includes \citet{mosser14} and \citet{vrard16} as 
training samples), followed by \citet{elsworth17}, \citet{bedding11}, and \citet{stello13}. Note that there are 713 stars without 
classifications, among which we label the 7 targets with $\numax>125\ \muHz$ as RGB stars, and the remaining 
706 stars as unclassified. For convenience, we provide all three solutions for every star. 
Some users may prefer to apply different corrections, and also the evolutionary stages of some stars may be revised in the future.

As shown in Figure \ref{figmassradius}, the \Dnu\ correction leads to overall lower mass estimates. Radii are less affected, since  
the radius scaling relation has less dependence on \Dnu\ (Equation \ref{radiusscaling}).

We can see from the mass histogram that our full sample covers a large stellar mass range, centered around 1.3 \msun\ and 
slightly skewed toward high-mass stars. The radius distribution sheds some light on the evolutionary stage. Red clump stars 
are expected to pile up around 11 \rsun\ due to their slower evolutionary rates compared to RGB stars. The sharp cutoff at the 
low-radius endpoint is associated with our sample selection, which does not include subgiants or main-sequence stars. As pointed 
out by \citet{kallinger10}, we expect to see a small excess near $R=$ 20.5 \rsun, made up of stars in the AGB clump phase 
\citep{cassisi01}. That excess is not readily apparent in our sample. Insets show close-ups of the distributions at high 
mass and radius. We checked the six most massive stars, with mass $>$4.5 \msun, and found that they   
indeed show relatively smaller large separations than more typical stars at the same \numax, resulting in 
larger seismically inferred masses. An investigation of the underlying physics by means of individual frequency modeling is 
in preparation.

We also provide surface gravities for over 16,000 stars derived from the scaling relation (Equation \ref{loggseiseq}). 
Surface gravity estimates from asteroseismic analysis are believed to be more precise ($\sim$~2\%, \citealt{hekker13}), 
compared to the time scale technique ($\sim$~4\%, \citealt{kallinger16}), the 8 hr ``flicker" method ($\sim$~25\%, \citealt{bastien13}), 
spectroscopy ($\sim$~50\%, \citealt{valenti05}), and photometric colors ($\sim$~100\%, \citealt{brown11}). 
Seismically derived log $g$ values can be used as constraints to lift the degeneracy when spectroscopically 
deriving effective temperature, surface gravity, and metallicity \citep{bruntt12,huber13}. Red clump stars are expected to 
lie at log $g$ $\sim$ 2.4 dex. Figure \ref{figmassradius} also indicates that our sample includes a significant number 
of low-luminosity red giants (log $g$ $\ga$ 2.6) and substantially fewer high-luminosity red giants (log $g$ $\la$ 2.3). 

\subsection{Uncertainties}
\label{uncertainty}
\begin{figure*}
\begin{center}
\includegraphics[width=\textwidth, height=16cm, keepaspectratio]{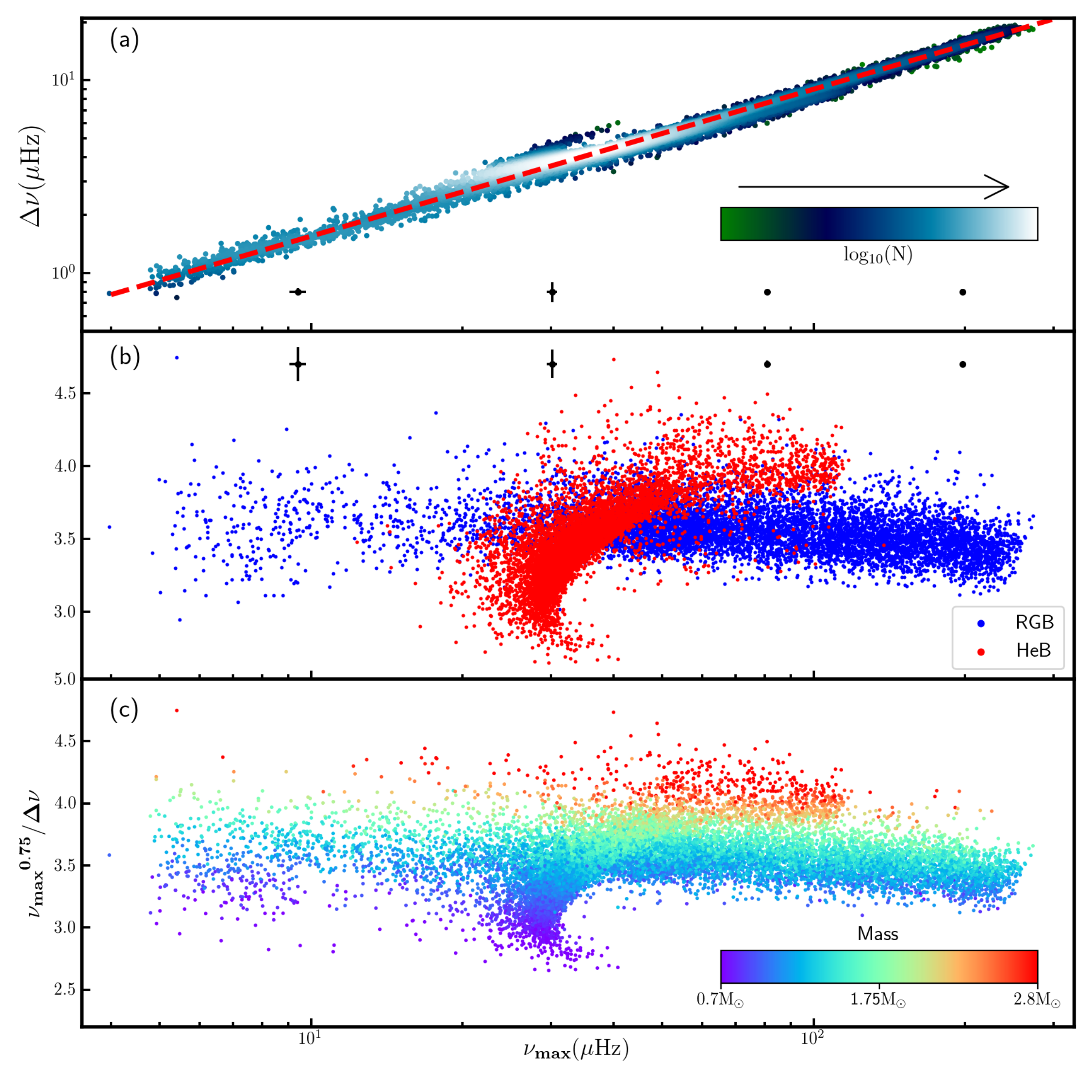}
\caption{\textbf{(a)} Power-law relation between \numax\ and \Dnu\, with number density of stars color-coded. The red dashed line 
is fitted using MCMC as $\Dnu = \alpha \cdot (\numax)^{\beta}$, where $\alpha = 0.267 \pm 0.002$, $\beta = 0.764 \pm 0.002$. 
\textbf{(b)} The scatter relative to the global trend color-coded with evolutionary phase. 
Stars without the phase classification, flagged as 0, are not shown here. Typical uncertainties are displayed. \textbf{(c)} Similar 
to panel (b) but color-coded with seismically inferred mass (see Section \ref{params}).}
\label{dnunumax}
\end{center}
\end{figure*}
The fractional uncertainties of seismic parameters and stellar fundamental properties are shown in Figure \ref{uncertainties}, 
based on full length end-of-mission long-cadence datasets. Considering output offset from different 
methods \citep{huber17}, we have added in quadrature 1.0\% and 0.5\% relative uncertainties in \numax\ and \Dnu\ to their formal 
uncertainties, respectively, only when deriving the uncertainties for mass, radius, and \logg. The correction factor, $f_{\Delta\nu}$, 
was fixed when determining the uncertainties of mass and radius using error propagation.

It can be seen from Figure \ref{uncertainties} that \numax\ and \Dnu\ can be measured more precisely in RGB stars (blue) 
than in HeB stars (red). This is mainly because HeB stars generally exhibit more complicated power spectra. This effect 
is propagated into the estimates of mass, radius, and \logg, which are derived from the scaling relations. 

The effective temperatures from \citet{mathur17} have four populations of distinct uncertainty distributions,  
from the KIC (\citealt{brown11}, $\sim$~4.0\%), the DR24 stellar properties catalog (\citealt{huber14}, $\sim$~4.0\%), 
the revised catalog of temperatures for long-cadence stars in the KIC (\citealt{pinsonneault12}, $\sim$~3.0\%), and the high-resolution 
spectroscopy from APOGEE DR12 (\citealt{pinsonneault14}, $\sim$~2.0\%, and \citealt{alam15}, $\sim$~1.7\%).

The uncertainty of the granulation power for RGB stars is bimodal, with a significant peak at higher 
uncertainties, while the HeB has a similar distribution to the RGB at lower uncertainties. These features are a result 
of increasing fractional uncertainties of granulation power with \numax\ and combined with the star number distribution as a 
function of \numax\ of the entire sample. For the subsample with $\numax < 30\ \muHz$, we found that RGB and HeB stars have 
very similar and single-peaked fractional uncertainty distributions that peak just below 0.1. As \numax\ increases, we see 
the right component gradually more clearly.

As expected, longer datasets enable us to determine global seismic parameters more precisely \citep{hekker12}. 
Thanks to the full-mission data sets used in our asteroseismic analysis (see Fig. \ref{figquarters}), we report precise 
determinations of global seismic parameters and stellar fundamental properties, with typical (median) precisions of 1.6\% in \numax, 
0.6\% in \Dnu, 4.7\% in oscillation amplitude, 8.6\% in granulation power, 8.8\% in width of power excess, 
7.8\% in mass, 2.9\% in radius, and 0.01 dex (or 0.4\%) in \logg. Considering only the 7839 stars with near full-mission 
data (time series longer than 1200 days), the uncertainty distributions shift slightly to lower values, while the overall distributions 
shapes look similar. The typical (median) fractional uncertainties are 1.4\% in \numax, 0.4\% in \Dnu, 4.6\% in oscillation amplitude, 
7.8\% in granulation power, 7.9\% in width of power excess, 6.9\% in mass, 2.5\% in radius, and 0.01 dex (or 0.3\%) in \logg.
The seismic and non-seismic parameters and their uncertainties are given in Table \ref{seisparatable} and \ref{stellarparatable}.

\subsection{Correlation between \numax\ and \Dnu}
\label{seisparams}
Figure \ref{dnunumax} displays the well-established power-law relation between \Dnu\ and \numax\ \citep{hekker09, stello09c, huber11b}. 
The red dashed line shown in Figure \ref{dnunumax}a was fitted using an MCMC method and is expressed as $\Dnu = \alpha \cdot (\numax)^{\beta}$, 
where $\alpha = 0.267 \pm 0.002$, $\beta = 0.764 \pm 0.002$. As noted previously in the literature, the power law is unable to perfectly 
describe the relation between \numax\ and \Dnu, especially in the common parameter space of RGB and HeB stars \citep{huber11b}. We can 
also see a concentration of HeB stars constituting a hook originating from around $\numax \simeq 30\ \muHz$, which we identify as the zero-age 
main-sequence for helium-core-burning.

In order to show this feature more clearly, we plotted $\numax^{0.75}/\Dnu$ as a function of \numax\ in Figure \ref{dnunumax}b and 
\ref{dnunumax}c, color-coded by evolutionary phase and stellar mass, respectively. The ordinate, $\numax^{0.75}/\Dnu$, has a mass 
dependence expressed as follows, by combining Equation \ref{numaxscale} and \ref{dnuscale}:
\begin{equation}
\frac {\left(\numax /\muHz \right)^{0.75}}{\Dnu /\muHz} \simeq \left(\frac{\rm M}{\rm M_{\sun}}\right)^{0.25}
\left(\frac{\rm T_{eff}}{\rm T _{eff, \sun}}\right)^{-0.375}.
\end{equation}
Since red giants cover a relatively small range of effective temperature, there is a very minor effective temperature influence on the 
distribution in Figure \ref{dnunumax}b and \ref{dnunumax}c. A pronounced feature in Figure \ref{dnunumax}b is the distinct distributions 
of HeB and RGB stars. The HeB stars form a sharp hook-shaped structure originating from \numax\ $\sim$30 \muHz\ (low-mass stars) and 
extending to \numax\ at $\sim$120 \muHz\ (high-mass stars). Note that there exists a sharp and extremely well-defined edge related to 
red clump stars. This is likely associated with the fact that all stars below roughly 2 \msun\ ignite helium in fully degenerate cores 
of very similar mass. This sharp edge also tells us that the scaling relations work well for the zero-age 
main-sequence of HeB stars.

In Figure \ref{dnunumax}c we note that some red clump stars have low-mass around or below 0.7 \msun. The lack of low-luminosity RGB stars 
(\numax\ $>\ \sim$40 \muHz) with such low-mass implies that there could be systematics in mass inferred with the scaling relations, or that 
at least some of those low-mass RGB stars undergo mass loss before reaching the HeB phase.  Assuming the latter is the case, the slightly 
less sharp edge toward the lowest mass HeB stars, at roughly $\numax^{0.75}/\Dnu \simeq 2.8$ and $\numax \simeq 30~\muHz$, could be a sign of 
``chaotic" variation in mass loss. In addition, the lack of low-mass stars ($\lesssim$ 0.7 \msun), marked by purple dots, at a higher \numax\ 
regime ($\gtrsim$ 40 \muHz) and the presence of the low-mass stars in the range \numax\ $\lesssim$ 20 \muHz, possibly suggest that the low 
mass stars with \numax\ $\lesssim$ 20 \muHz\ are in the AGB phase and have experienced mass loss.

\subsection{Seismic H-R diagram}
\begin{figure}
\begin{center}
\includegraphics[width=\columnwidth]{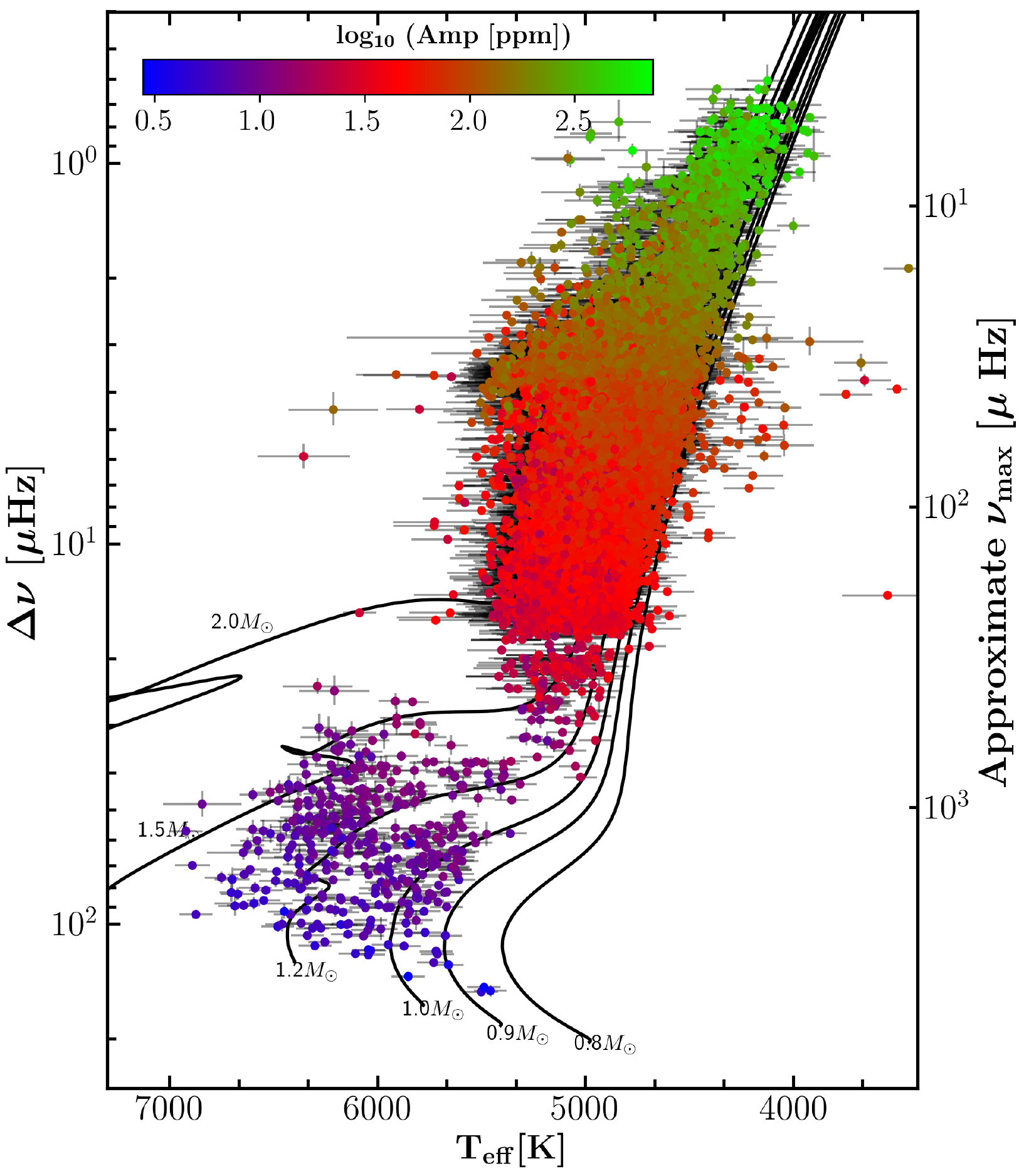}
\caption{Seismic H-R diagram (\Dnu\ vs. \teff). The estimates of \Dnu\ are from this work for red giants, 
except for super-Nyquist red giants from \citet{yu16}. The values of \Dnu\ for main-sequence and subgiant stars are collected from 
\citet{huber11b}. Approximate \numax\ values are shown on the right axis. Logarithmic oscillation amplitude per radial mode is color-coded. 
The solid lines show solar-metallicity evolutionary tracks, with mass labeled. Temperatures are adopted from \citet{mathur17}.}
\label{seishr}
\end{center}
\end{figure}
Figure \ref{seishr} shows a seismic H-R diagram for the largest sample of \kep\ solar-like oscillators so far, with \Dnu\ 
being measured from this work for red giants, except for super-Nyquist red giants from \citet{yu16}, and from \citet{huber11b} for main-sequence 
and subgiant stars. We select \Dnu\ rather than \numax\ to illustrate the seismic H-R diagram, because \Dnu\ can be more accurately measured for stars 
oscillating around the Nyquist frequency. Red giants oscillate with amplitudes ranging from a few tens to thousands of parts per million, as 
shown in the color. The characteristic oscillation timescales vary from hours up to days, as indicated by \numax\ plotted on the right. A 
few outliers are present, mainly due to the poorly determined temperatures for those stars \citep{mathur17}. The large uncertainties 
of the temperatures blur the distributions of red clump stars, making it difficult to distinguish from RGB stars. We observe a sharp edge 
lying at \numax\ $\simeq\ 275\ \muHz$ (\Dnu\ $\simeq\ 19.2\ \muHz$), corresponding to the upper limit of \numax\ estimates in this work. Solar-like 
oscillations with \numax\ greater than the long-cadence Nyquist frequency are generally detected with short-cadence data 
\citep{gilliland10b}. \citet{murphy13} and \citet{chaplin14}, however, pointed out that it remains possible to detect these oscillation using long-cadence 
data. \citet{yu16} subsequently identified 98 stars oscillating in the super-Nyquist frequency regime, up to 387 μHz. 
These super-Nyquist red giants are also plotted in Figure \ref{seishr}.

\begin{figure*}[htp]
\begin{center}
\includegraphics[width=\textwidth, height=12.8cm, keepaspectratio]{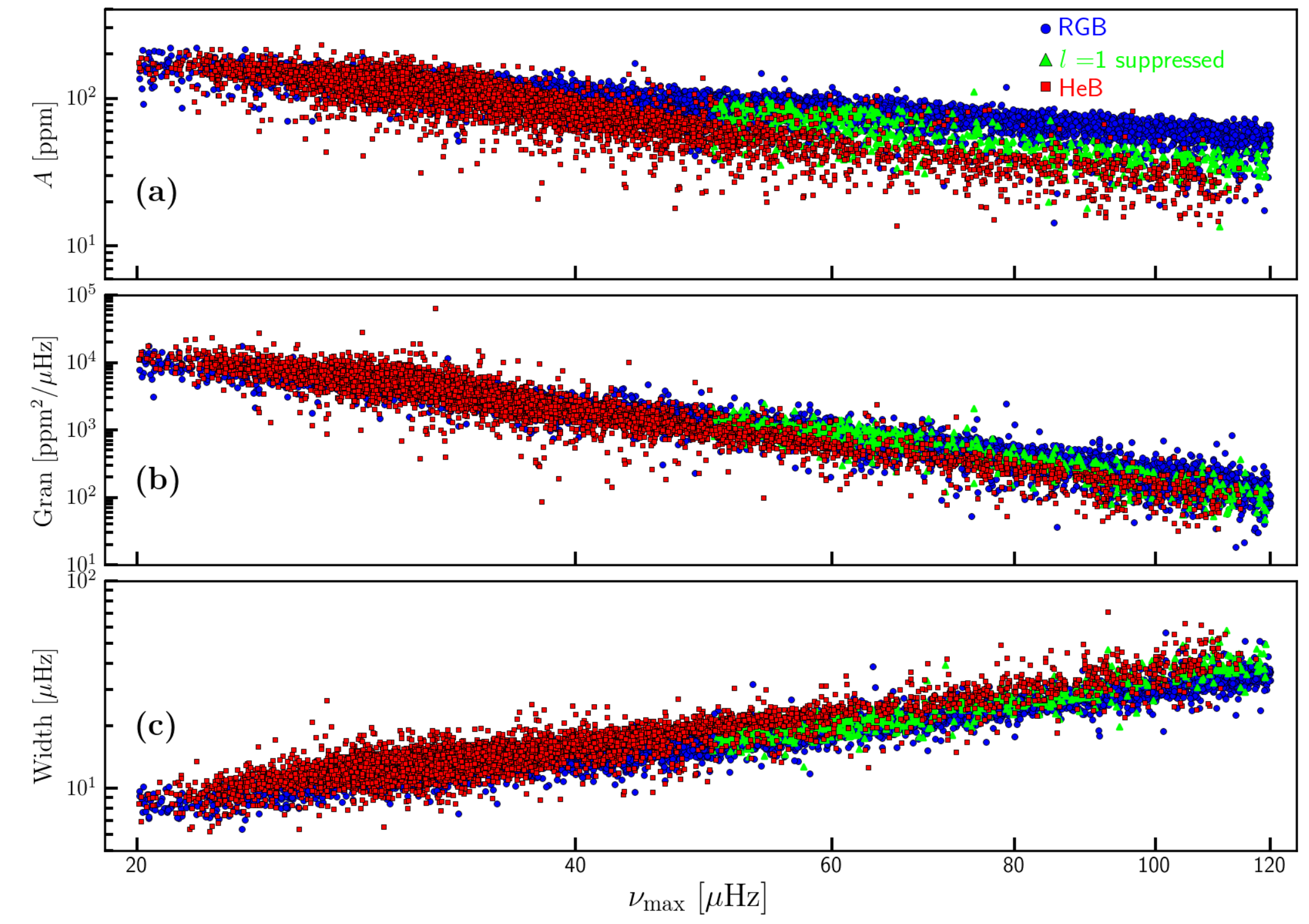}
\caption{Distributions of \textbf{(a)} oscillation amplitude, \textbf{(b)} granulation power measured at \numax, 
and \textbf{(c)} width of power excess for RGB (blue circles) and HeB (red squares) stars. The green triangles show 
dipole-mode suppressed oscillators \citep{stello16a}.}
\label{rgbheb}
\end{center}
\end{figure*}

\section{Mass and Metallicity effects on power excess parameters}
\subsection{Power excess difference in RGB and HeB stars}
Figure \ref{rgbheb} shows that RGB and HeB stars follow different distributions of oscillation amplitude, granulation power, 
and the width of power excess (see Section \ref{params} for the evolutionary phase classification). The differences are negligible 
for low \numax\ red clump stars (\numax\ $\simeq30$ \muHz), but gradually become substantial for higher \numax\ stars (\numax\ 
$\simeq60$ \muHz), and significant for secondary clump stars (HeB stars that did not undergo a helium flash, due to the 
non-degenerate helium core). We confirm that secondary clump stars have lower oscillation amplitudes and granulation power, and 
broader power excesses compared to RGB stars at a given \numax\ \citep{mosser12b}. We have tested and found that the offsets of both 
oscillation amplitude and granulation power between RGB and clump stars, in the range $50\ \muHz<\numax<120\ \muHz$, are mainly 
due to the difference in stellar mass between the two populations, followed by luminosity and temperature, using the formulas 
fitted by \citet{huber11b}. The oscillation amplitude and granulation power show tight correlations, and both increase when the 
star evolves up the RGB \citep{kjeldsen11, huber11b, kallinger14}. 

\begin{figure*}
\begin{center}
\includegraphics[width=\textwidth, height=12cm, keepaspectratio]{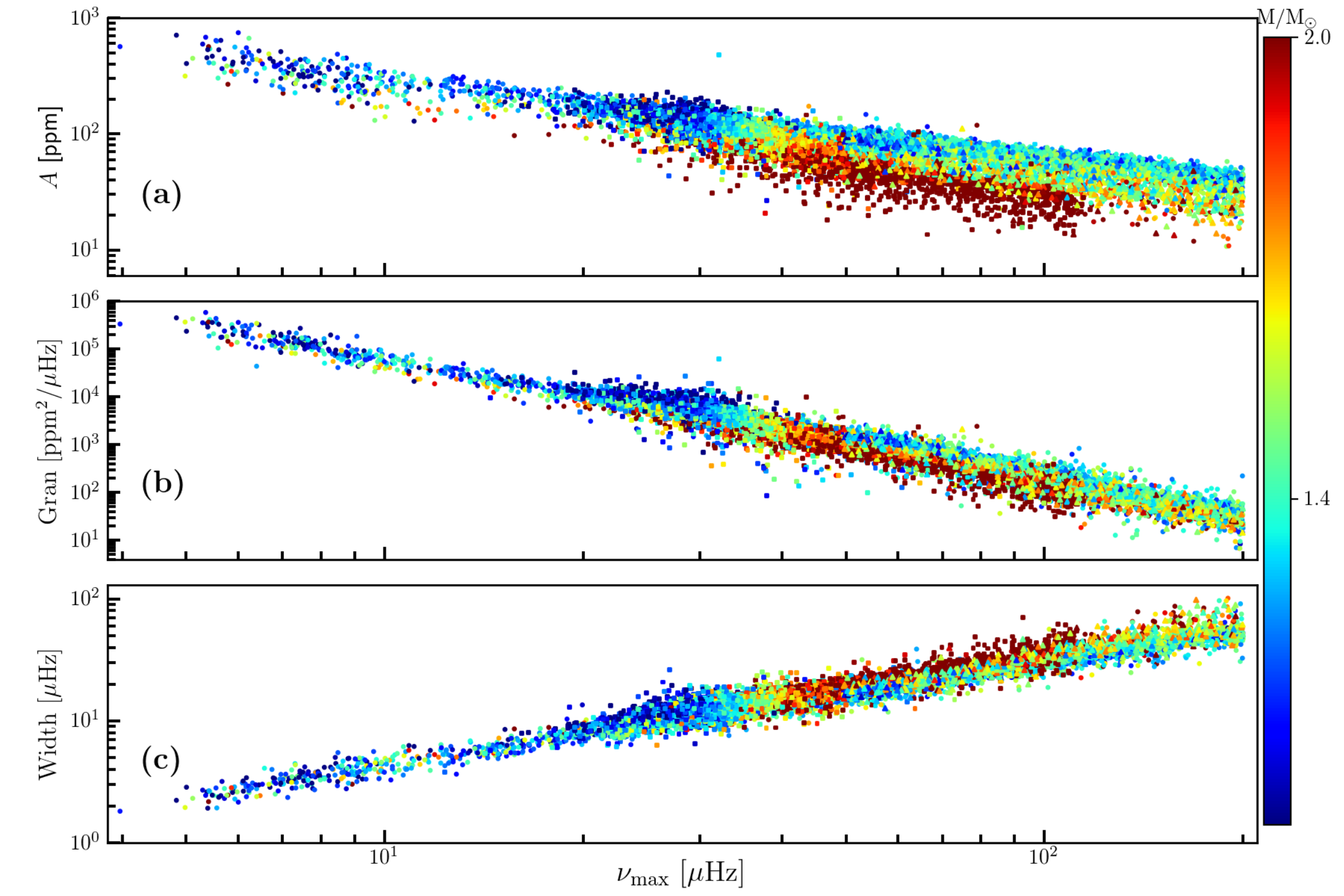}
\caption{The distributions of \textbf{(a)} oscillation amplitude, \textbf{(b)} granulation power measured at \numax, 
and \textbf{(c)} width of power excess for all red giants, color-coded by the seismic mass, which is truncated to display 
the mass effect. The measurements at \numax\ greater than 200 \muHz\ are not shown because of the difficulty in 
modeling spectrum background when power excess approaches the Nyquist frequency. The symbols have the same meaning as in Figure \ref{rgbheb}}
\label{masseffect}
\end{center}
\end{figure*}

\begin{figure*}[t]
\begin{center}
\includegraphics[width=\textwidth, keepaspectratio]{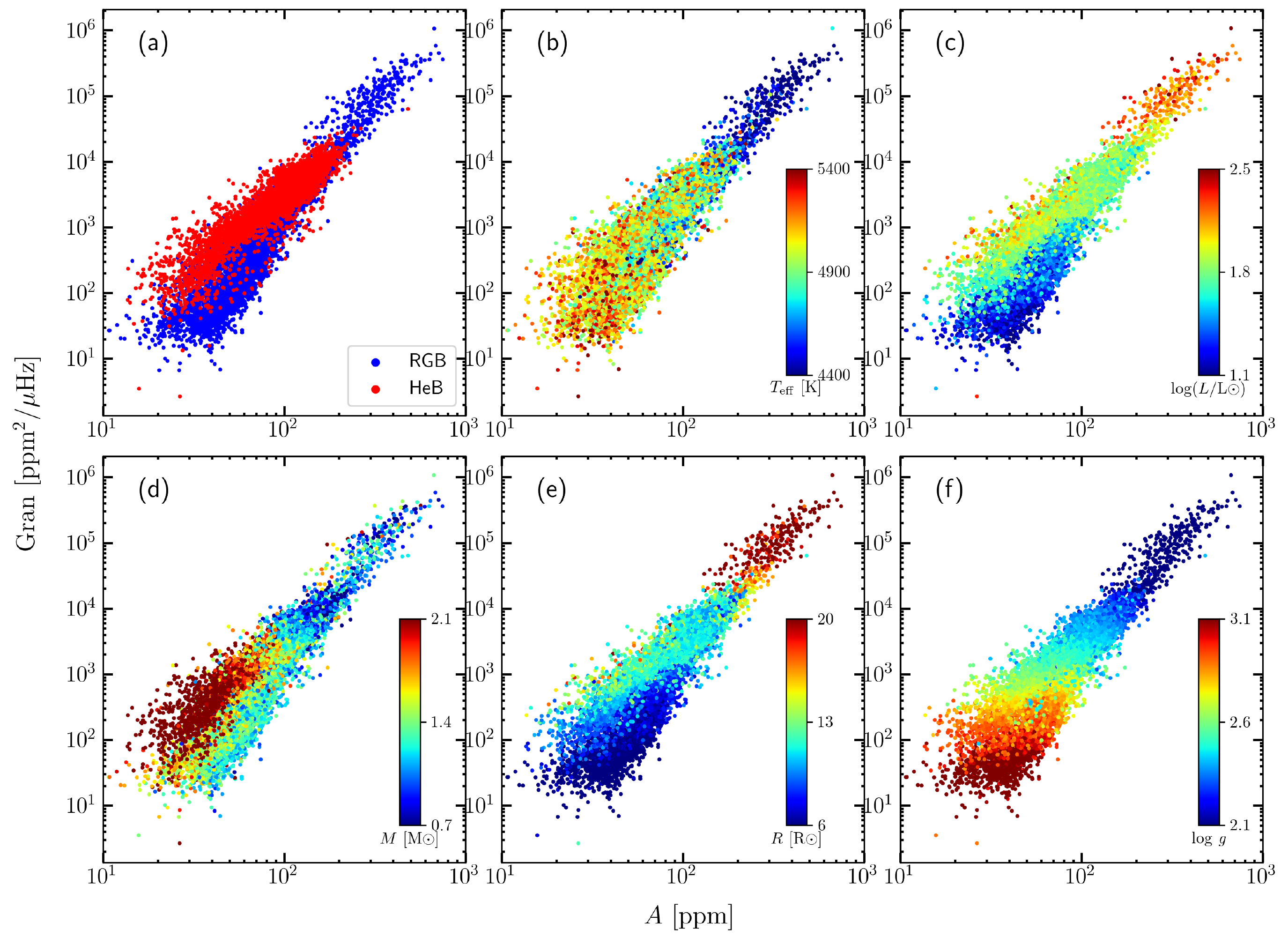}\\
\caption{Relation of oscillation amplitude and granulation power, color-coded by \textbf{(a)} evolutionary phase, 
\textbf{(b)} effective temperature, \textbf{(c)} luminosity, \textbf{(d)} mass, \textbf{(e)} radius, and \textbf{(f)} \logg. 
Stars with $\numax>200\ \muHz$ are excluded.}
\label{oscgra}
\end{center}
\end{figure*}

\begin{figure*}[t]
\begin{center}
\includegraphics[width=\textwidth, height=9.cm, keepaspectratio]{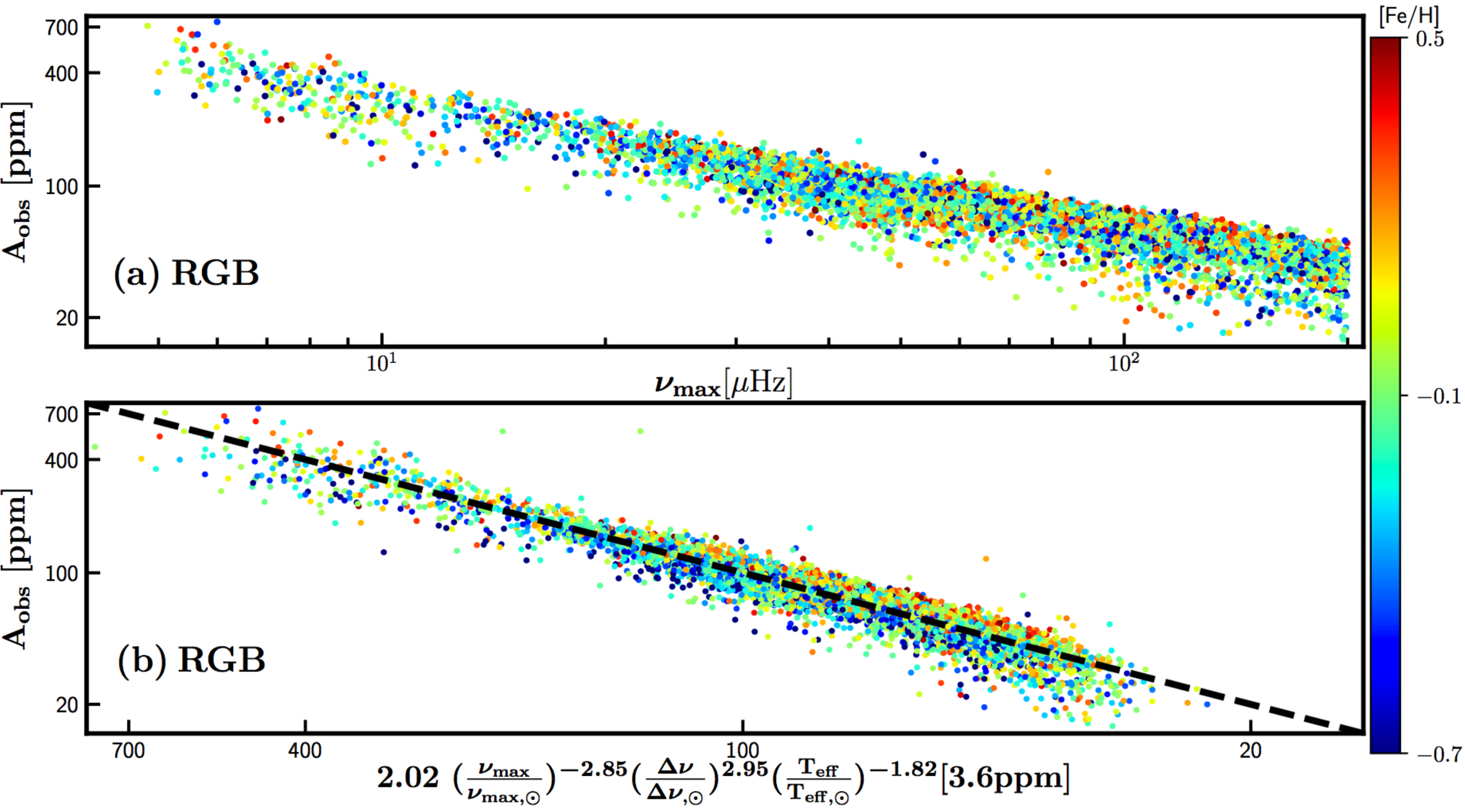}\\
\includegraphics[width=\textwidth, height=9.cm, keepaspectratio]{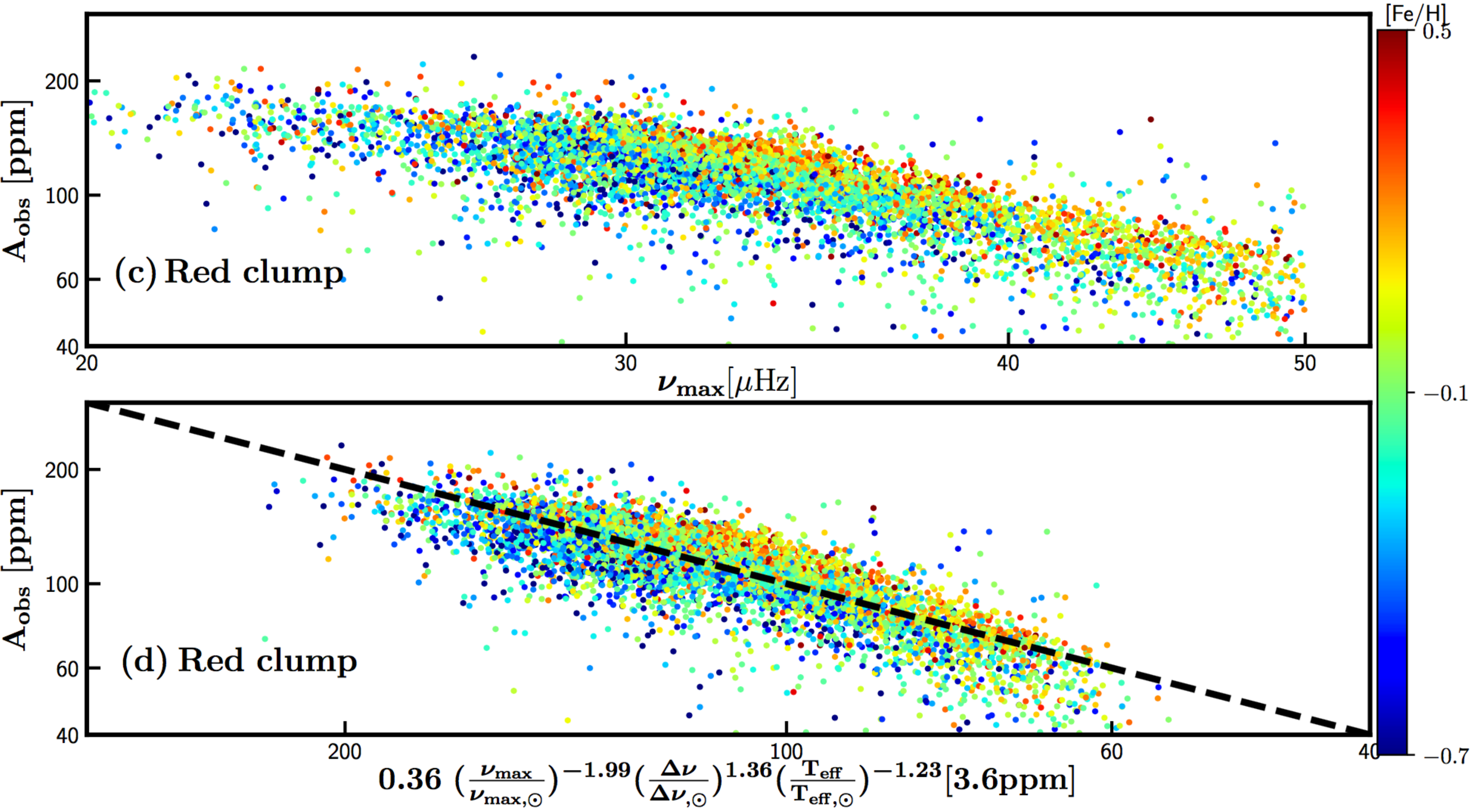}\\
\caption{Metallicity influence on the oscillation amplitude of RGB and red clump stars. \textbf{(a)} Observed 
oscillation amplitude is plotted against \numax, color-coded by metallicity. \textbf{(b)} The observed 
oscillation amplitude is compared to its calculated counterpart using the model of equation \ref{massscaling}. 
Panels \textbf{(c)} and \textbf{(d)} are similar to \textbf{(a)} and \textbf{(b)}, but are for red clump stars, separated 
from secondary clump stars with a threshold of \numax=50 \muHz. The black dashed line shows the one-to-one 
relation. Stars with $\numax>200\ \muHz$ have been excluded due to the difficulty of the modeling granulation 
background.}
\label{rgbfeh}
\end{center}
\end{figure*}

We checked whether the smoothing process is responsible for the broader power excesses of secondary clump stars. One 
might suspect that the broader power excess of a secondary clump star could artificially arise from the smoothing 
process applied to measure it. To test this, we selected 20 RGB and secondary clump stars, 
including stars with extreme width values, with the same \numax\ values in the range $60\ \muHz<\numax<120\ \muHz$ 
and checked individual original power spectra without smoothing. We found the power excess of the secondary clump 
stars to be intrinsically broader than the RGB stars. Therefore, the difference of power excess width in Figure \ref{rgbheb}c 
is not a measurement bias. The broader power excess might reduce the precision when 
measuring \numax, which will propagate into the seismic determinations of mass and radius.

The measured amplitude of dipole-mode suppressed RGB stars \citep{stello16a}, as shown in Figure \ref{rgbheb}a, is 
smaller than that of normal RGB stars. This is a simple consequence of the lack of power in the dipole modes, which 
constitute about half the total power \citep{stello16a}. We find that the suppressed stars show $\sim$\ 9\% less 
granulation power than normal RGB stars, as shown in Figure \ref{rgbheb}b. This offset is presumably caused by the mass 
difference between the dipole-mode suppressed stars and the normal RGB stars. To understand this, we recall that (1) 
dipole-mode suppressed stars show a mass distribution shifted to larger masses compared to normal RGB stars 
\citep[see Figure \ref{masseffect}a, and Figure 2 in ][]{stello16a}. (2) Granulation power is 
a decreasing function of stellar mass, as predicted by \citet{kjeldsen11}. (3) The granulation power of dipole-mode  
suppressed stars follows the same relations as normal red giants \citep{garcia14b}.

\subsection{Mass effect}
Many efforts have been made to investigate the dependence of oscillation amplitude and granulation power on stellar fundamental 
properties in field stars and open clusters \citep[e.g., ][]{kjeldsen08, kjeldsen11, stello11, huber11b, mathur11b, mosser12b, 
corsaro13, kallinger14, corsaro17b}. The empirical relations as proposed by \citet{kjeldsen08} and \citet{kjeldsen11} predict that 
oscillation amplitude and granulation power depend on the luminosity, mass, and temperature. In this section we study the mass 
influence on oscillation and granulation using our sample of over 16,000 oscillating red giants.

As shown in Figure \ref{masseffect}a, the oscillation amplitude is a decreasing function of \numax, ranging from $\sim$ 
600 ppm at \numax $\simeq$ 5 \muHz\ down to $\sim$ 20 ppm at \numax $\simeq$ 200 \muHz. We observe an extremely 
sharp upper boundary, which might be related to excitation and damping of oscillation modes. We note 
that there are a number of stars with low amplitudes along the bottom of the global trend. We have checked those stars and 
found that some have low S/N power spectra, with \kep\ time series that are only a few quarters long. The low-amplitude stars 
could also be contaminated by nearby stars or diluted by their companions in binary systems \citep{ziegler17, schonhut17} 
or be exotic stars. Figure \ref{masseffect}a also shows that high-mass HeB stars have overall lower amplitudes compared to 
RGB stars at a given \numax. A clear mass gradient is present in RGB stars as well. 

The granulation power determined at \numax\ has a dependence on stellar mass, as shown in Figure \ref{masseffect}b. Our results are 
in qualitative agreement with the predictions given by \citet{kjeldsen11}, where higher-mass stars are expected to have lower 
oscillation amplitude and granulation power \citep{mosser12b}. 

Red giants show progressively narrower power excess when evolving toward the tip of the RGB. Only approximately three 
orders of modes can be detected in high-luminosity red giants with \numax\ $\simeq$ 5 \muHz\ and seven orders of modes 
in low-luminosity red giants \citep{stello14, corsaro15a}. We confirm that the width of power excess is an increasing function 
of stellar mass as shown in Figure \ref{masseffect}(c) \citep[see also][]{mosser12b}. Higher-mass stars with wider 
power excess and lower oscillation amplitude imply that the total oscillation power tends to be almost conserved among the 
stars of different mass \citep{kallinger14}.

As shown in Figure \ref{masseffect}, granulation power correlates with oscillation amplitude. To see this more 
clearly, we plotted oscillation amplitude against granulation power in Figure \ref{oscgra}a, color-coded by the evolutionary phase (RGB 
or HeB). The correlations are similar but distinct for RGB and HeB stars \citep{kallinger14}. The two populations following different 
distributions are mainly caused by their different stellar fundamental properties: luminosity, mass, and temperature (thus radius and 
\logg), as we can see from Figures \ref{oscgra}b-f. The temperature effect stands out among RGB stars but is not clear between RGB and 
HeB at a given oscillation amplitude, because of their similar temperatures. Our results are qualitatively consistent with the predictions 
by \citet{kjeldsen11}.

\subsection{Metallicity effect}
Figure \ref{rgbfeh} shows the \numax-amplitude relation color-coded by metallicity.
We observe that in Figure \ref{rgbfeh}a the metallicity effect is not visually striking, which is due to the fact that 
oscillation amplitudes have a stronger dependence on luminosity, mass, and temperature \citep{kjeldsen11}, blurring the possible 
metallicity influence. In this section we attempt to investigate the metallicity effect on oscillation and granulation. Instead of  
fitting to the luminosity and mass inferred from the scaling relations with possible systematics, we fit the oscillation amplitude as 
a function of the observables, \numax, \Dnu, and temperature, as
\begin{equation}
{ \frac{A}{\rm A_{\odot}} = \alpha \left(\frac{\nu_{\rm max}} {\nu_{\rm max, \odot}}\right)^\beta \left(\frac{\Delta\nu}
{\Delta\nu_{,\odot}}\right)^\gamma \left(\frac{T_{\rm eff}}{\rm T_{eff,\odot}}\right)^\delta.}
\label{massscaling}
\end{equation}
Here, $\alpha$ is a scaling factor introduced so that our model does not have to pass through the solar reference point ($\rm{A_{\odot}}$ 
= 3.6 ppm). Equation \ref{massscaling} was converted to a logarithmic scale when implementing the fit. We attempted to study the metallicity 
influence on RGB and HeB stars separately, because the two populations have different distributions, as shown in Figure \ref{rgbheb}a and 
Figure \ref{masseffect}a. In this work, metallicity estimates are adopted from \citet{mathur17}. We excluded all the stars with \numax\ 
$>$ 200 \muHz\ because the backgrounds of their power spectra are hard to model due to the oscillations being close to the Nyquist frequency. 

\begin{figure*}
\begin{center}
\includegraphics[width=\textwidth, height=10cm, keepaspectratio]{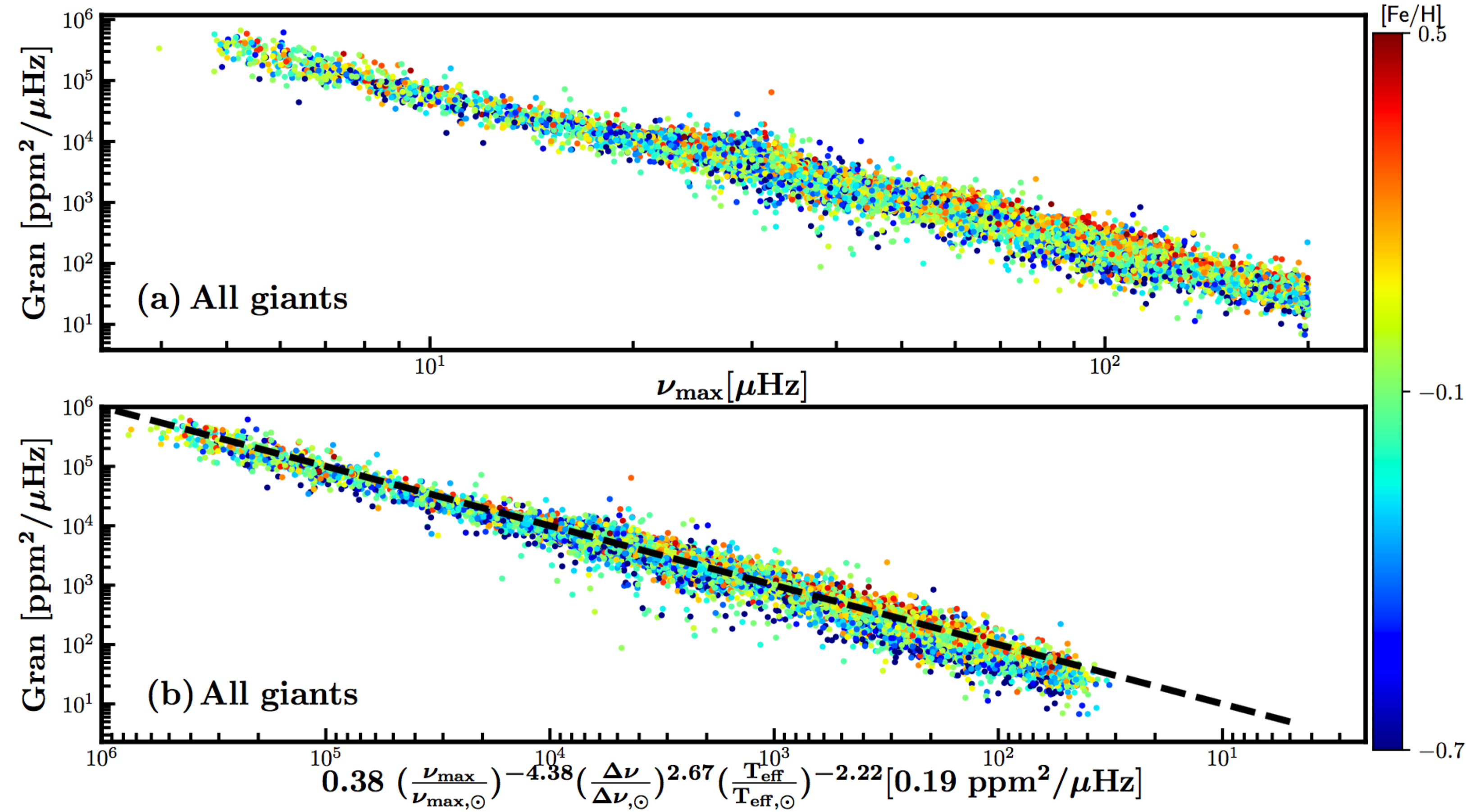}
\caption{Similar to Figure \ref{rgbfeh} but for granulation power of all the red giants.}
\label{gramfeh}
\end{center}
\end{figure*}

By removing the underlying contributions from luminosity, mass, and temperature \citep{kjeldsen11}, the significant metallicity influence 
stands out clearly, as shown in Figure \ref{rgbfeh}b, where metal-rich stars oscillate with larger amplitudes than metal-poor stars. The 
fitted values and the corresponding uncertainties of the parameters are shown in Table \ref{fitpara}. The fitted scaling factor \textcolor{
blue}{\textbf{$\alpha=2.02$}} differs from unity, consistent with the results by \citet{corsaro17b}.

We repeated the fit on the HeB stars. The red clump and secondary clump stars in our sample differ significantly in metallicity. 
The former covers a much larger metallicity range than the latter, which are overall more metal rich. Considering our primary goal 
of investigating metallicity effects, we excluded secondary clump stars using a threshold $\numax>50\ \muHz$. Figure 
\ref{rgbfeh}d shows that the metallicity influence also exists for red clump stars, where metal-rich stars again oscillate with higher 
amplitudes at given \numax, \Dnu, and \teff. 

We note that the oscillation amplitudes of RGB and red clump stars have different levels of dependence on \numax, \Dnu, and temperature, 
as the two relations show in Figure~\ref{rgbfeh}. The influences of \numax, \Dnu, and temperature on RGB stars are more significant than those 
on red clump stars, as revealed by the globally larger absolute exponents of the relation for RGB stars versus those for red clump stars 
(see Table \ref{fitpara}). The different influence is because the \numax, \Dnu, and temperatures of RGB stars 
vary more significantly than those for HeB stars.

\begin{table}[b]
\begin{small}
\begin{center}
\caption{Fitted model parameters of oscillation amplitude and granulation power}
\resizebox{\columnwidth}{!}{\begin{tabular}{llllll}
\hline
\hline
Parameters  & phase  & $\alpha$ & $\beta$ & $\gamma$ &  $\delta$ \\
\hline
\multirow{2}{*}{Oscil. amp.} & RGB &  2.02$\pm$0.02  & -2.85$\pm$0.04  & 2.95$\pm$0.05  & -1.82$\pm$0.05\\
\cline{2-6}
& Red clump                        &  0.36$\pm$0.03  & -1.99$\pm$0.02  & 1.36$\pm$0.04  &-1.23$\pm$0.05\\
\hline
Gran. power           & Red giants &  0.38$\pm$0.01  & -4.38$\pm$0.01  & 2.67$\pm$0.03  & -2.22$\pm$0.05\\
\hline
\end{tabular}}
\label{fitpara}
\end{center}
\flushleft 
Notes: The fitted values and uncertainties of the parameters, namely $\alpha, \beta, \gamma, \delta$, as defined in Equation 
\ref{massscaling}. The oscillation amplitudes of RGB and red clump stars were fitted separately, while the granulation power 
of all red giants was fitted to the entire sample.
\end{small}
\end{table}

Since the intensity fluctuation caused by granulation is related to the contrast between dark and bright regions of 
granules, the opacity and limb-darkening should have a strong effect on granulation power. Thus,   
metallicity in turn might influence granulation power as well. Figure \ref{gramfeh} displays the metallicity 
effect on granulation power measured at \numax, using the same model as shown in Equation \ref{massscaling}. From Figure 
\ref{gramfeh}a, we observe that the metallicity effect is visible in the \numax\ range covered by secondary clump stars. With 
the attempt to remove underlying contributions from luminosity, mass, and temperature, Figure \ref{gramfeh}b indicates 
that the granulation power depends on metallicity, where metal-rich stars have larger granulation power than 
metal-poor stars.

Our results reveals that the effect of metallicity on the granulation power of field red giants are in agreement with the arguments given 
by \citet{corsaro17b}, who found that metallicity causes a statistically significant variation in the amplitude of the granulation 
activity of stars in the open clusters NGC 6791, NGC 6819, and NGC 6811. \citet{collet06} performed 3D hydrodynamical 
simulations of red giants with [Fe/H] from -3.0 through 0.0, and found that more metal-rich stars have larger granules (see their 
Figure 4) due to increased opacity. The increased horizontal size of a granule suggests metal-rich stars are expected to have 
greater granulation power, which is qualitatively consistent with our results. To understand this, we recall that a convection cell 
is usually assumed to travel a vertical distance proportional to the pressure scale height, $H_{\rm{p}}$, at a speed scaling with 
the sound speed, $c_{\rm{s}}$, thus the the characteristic timescale of a granule can be expressed as $\tau_{\rm{gran}} \propto 
H_{\rm{p}}/c_{s} \propto \left(T_{\rm{eff}}/g)\right/\sqrt{T_{\rm{eff}}} \propto 1/\nu_{\rm_{max}}$. The assumption of the 
horizontal size of a granule, $d$, proportional to the pressure scale height yields $d \propto H_{\rm{p}} \propto T_{\rm{eff}}/g 
\propto \sqrt{\rm{T_{eff}}}/\nu_{\rm{max}}$. The proportionality of granulation power to $\sigma^2 \tau_{\rm{gran}}$, with 
$\sigma$ being the rms intensity fluctuation and varying much less than $\tau_{\rm{gran}}$, gives the granulation power 
${\rm{Gran}} \propto d /\sqrt{T_{\rm{teff}}}$, hence, larger granules are expected to have larger granulation power 
\citep[for more detail see][]{kjeldsen11, mathur11b}.

\begin{figure}
\begin{center}
\resizebox{\columnwidth}{!}{\includegraphics{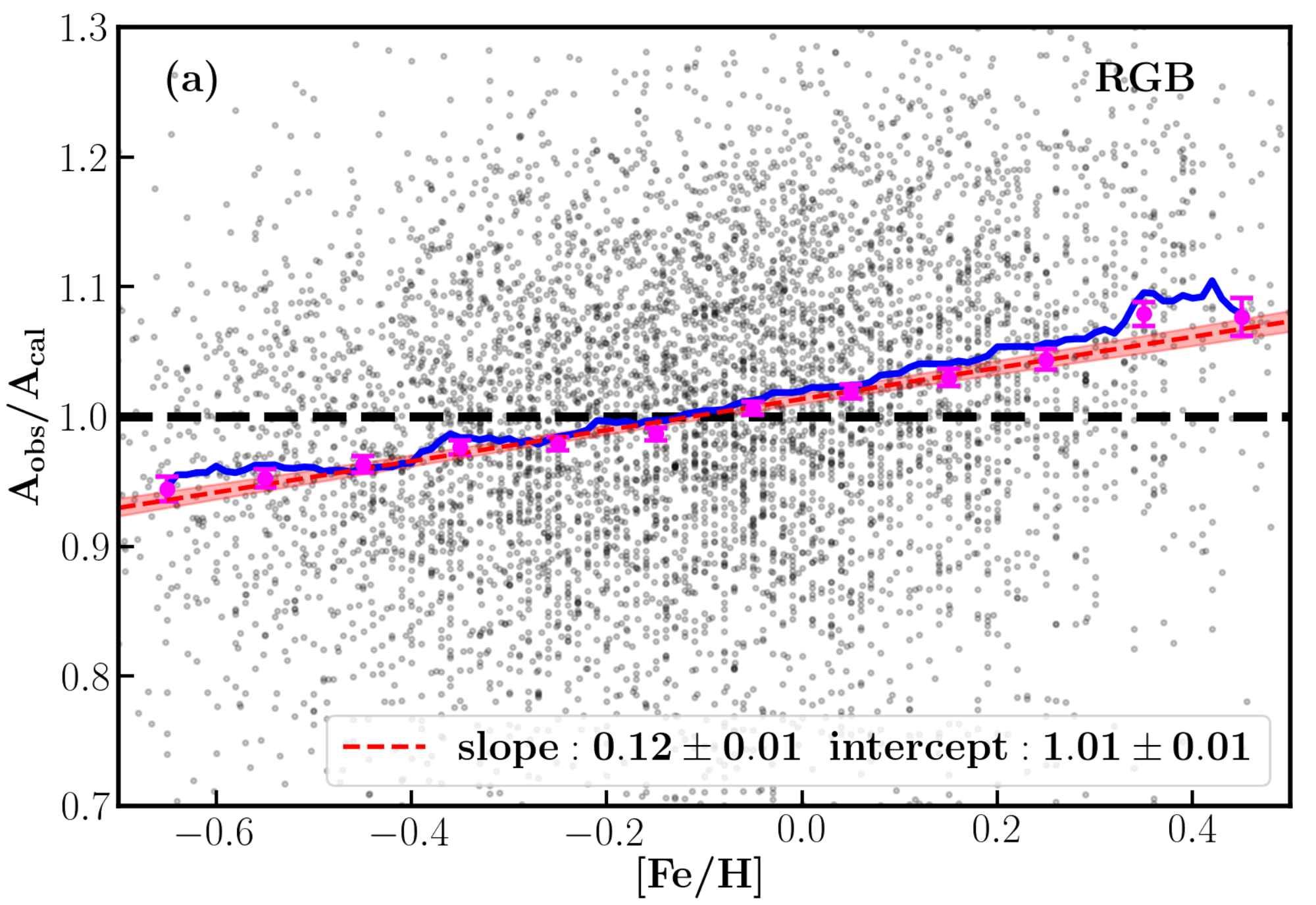}}\\
\resizebox{\columnwidth}{!}{\includegraphics{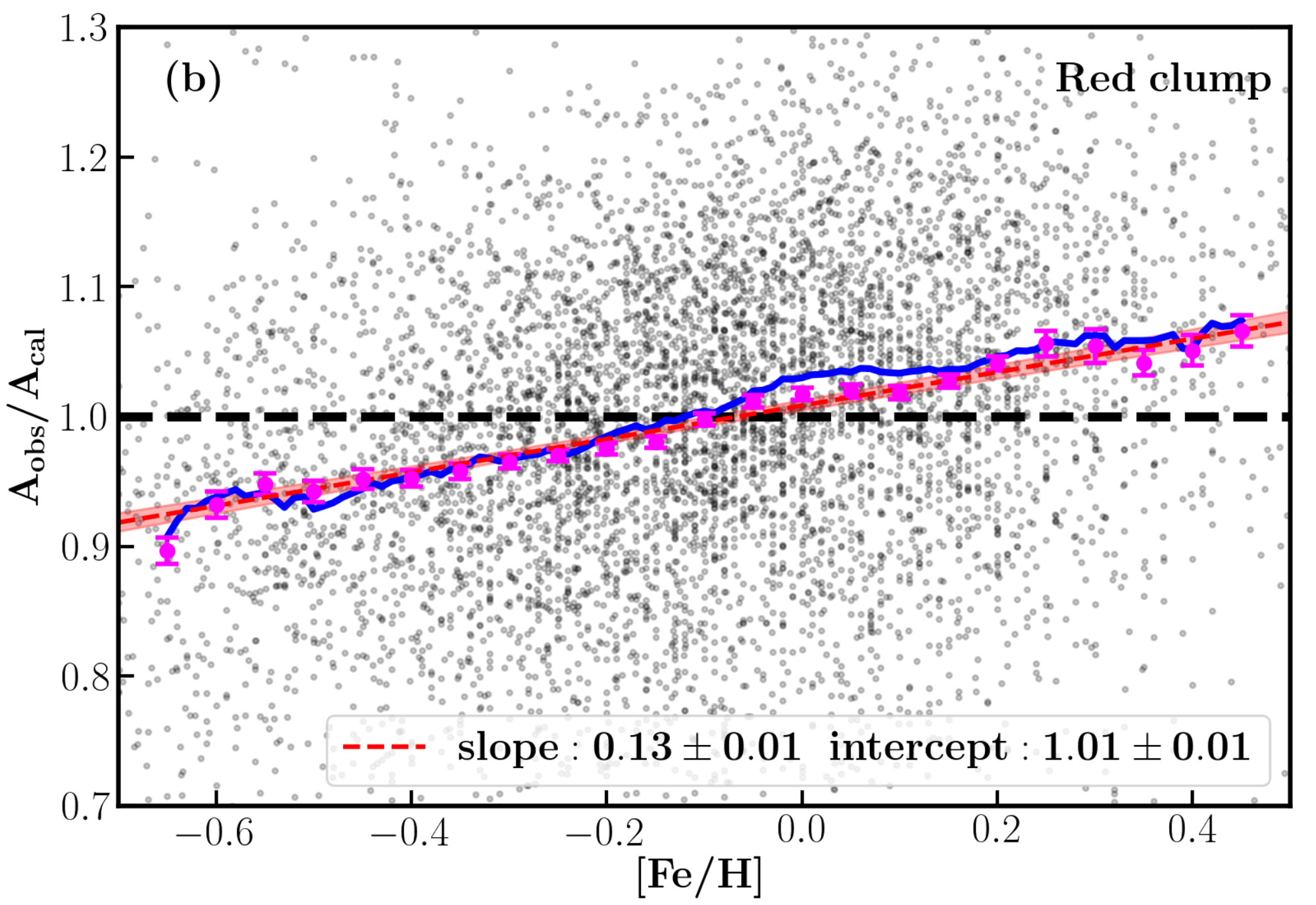}}\\
\resizebox{\columnwidth}{!}{\includegraphics{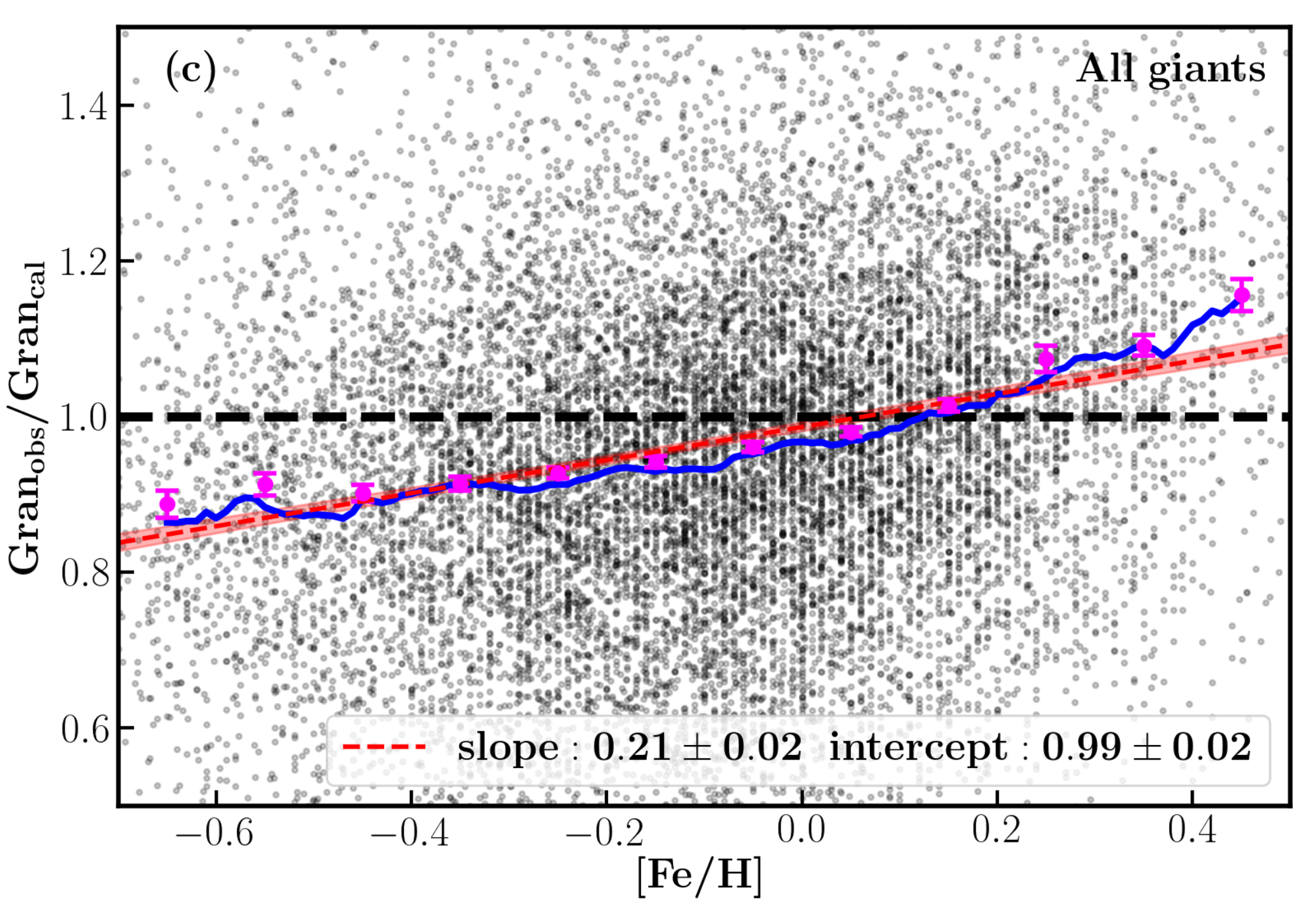}}\\
\caption{\textbf{(a)} The residual of the observed and predicted oscillation amplitude for RGB stars and \textbf{(b)} for red clump stars, 
and \textbf{(c)} the residual of the observed and predicted granulation power of red giants in the whole sample, as a function of [Fe/H]. 
A threshold of $\numax>50\ \muHz$ is used to separate red clump stars from secondary clump stars. In each panel, 
the blue line denotes the 50th percentiles of the residuals. The magenta points mark the mean residuals with error 
bars in each 1.0 dex wide bin. A linear fit to individual data points is shown by the red dashed line, with 95\% confidence 
intervals in each box. The slopes and intercepts are indicated.}
\label{fig:fehresidual}
\end{center}
\end{figure}

\begin{figure}
\begin{center}
\resizebox{\columnwidth}{!}{\includegraphics{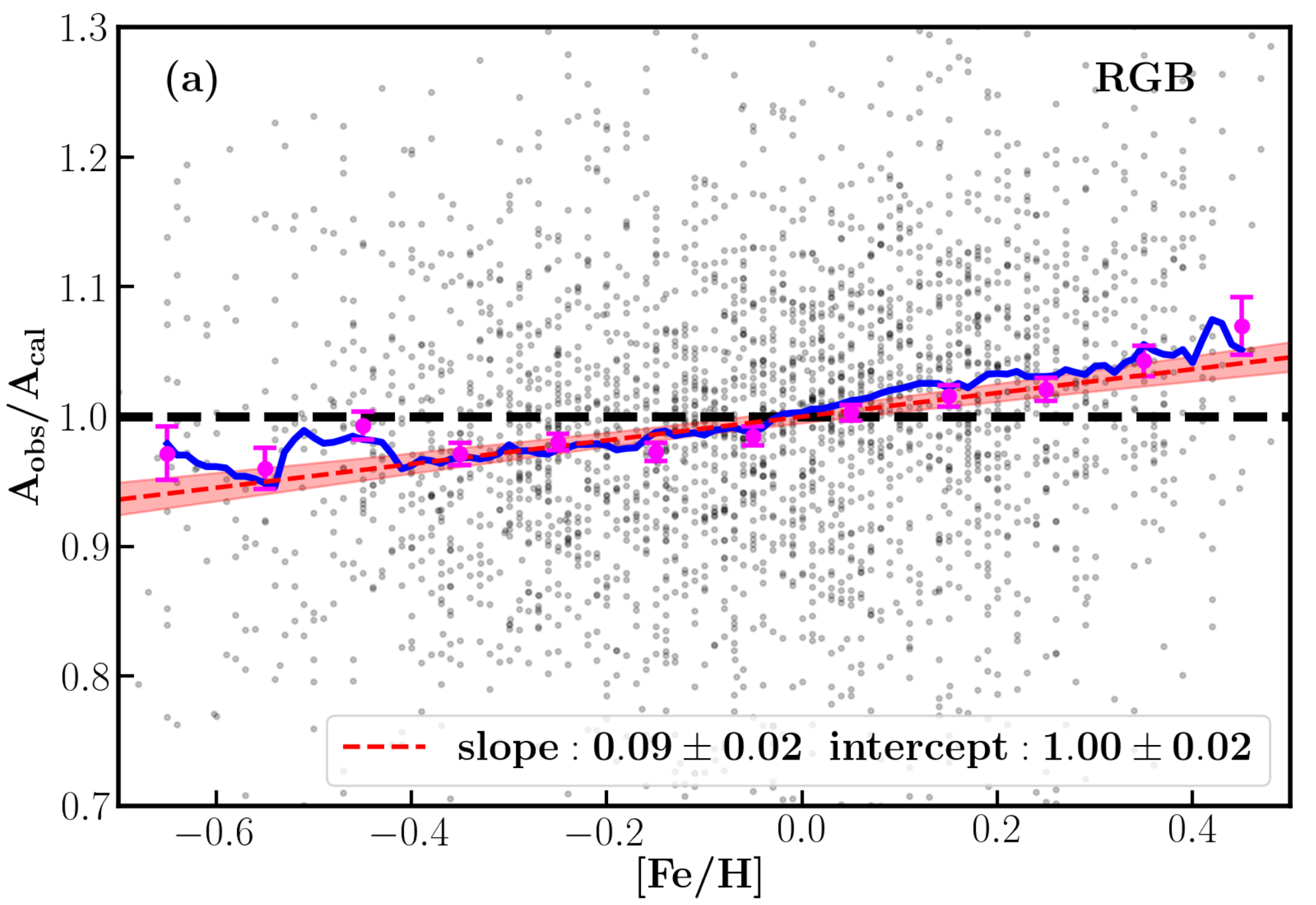}}\\
\resizebox{\columnwidth}{!}{\includegraphics{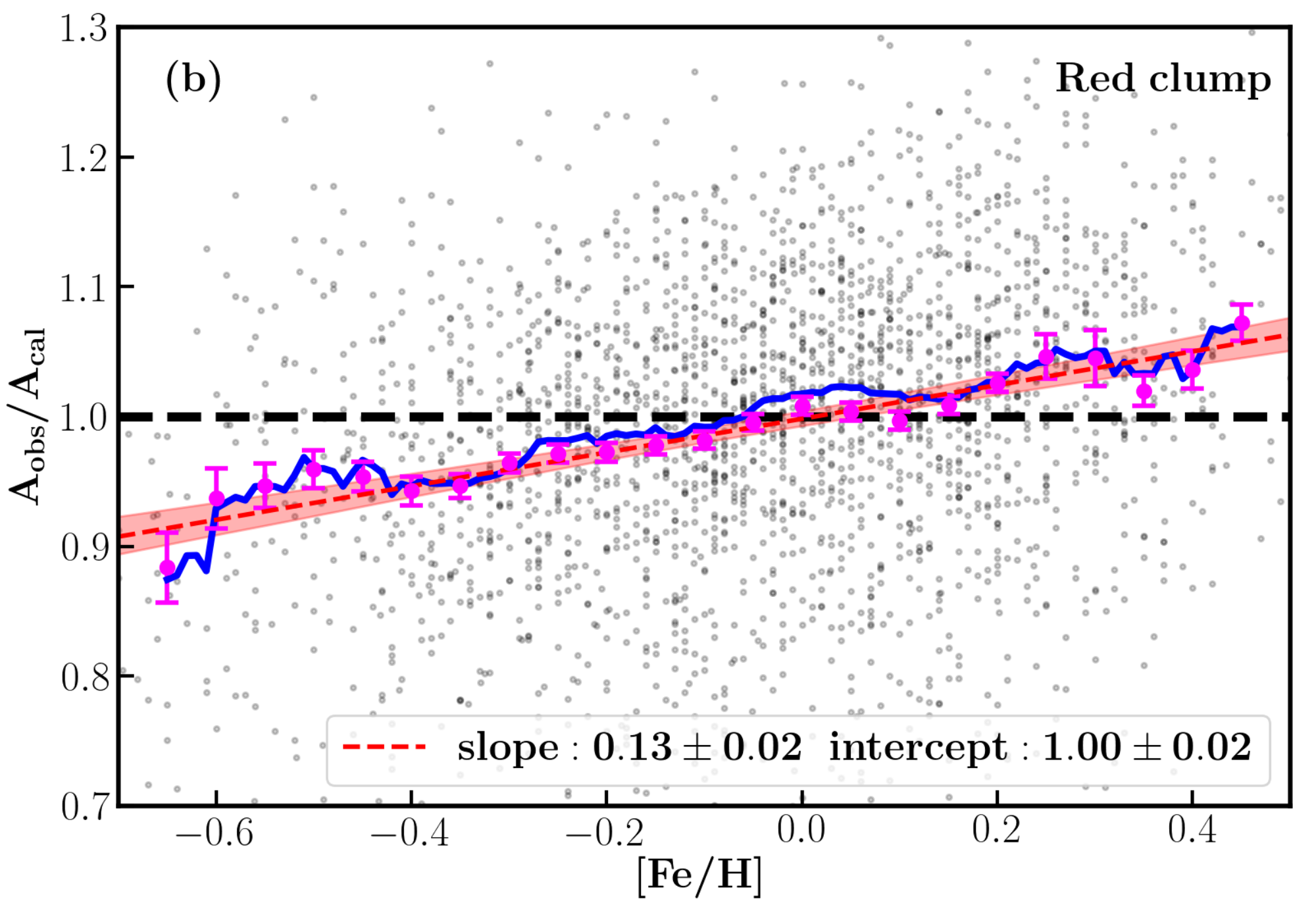}}\\
\resizebox{\columnwidth}{!}{\includegraphics{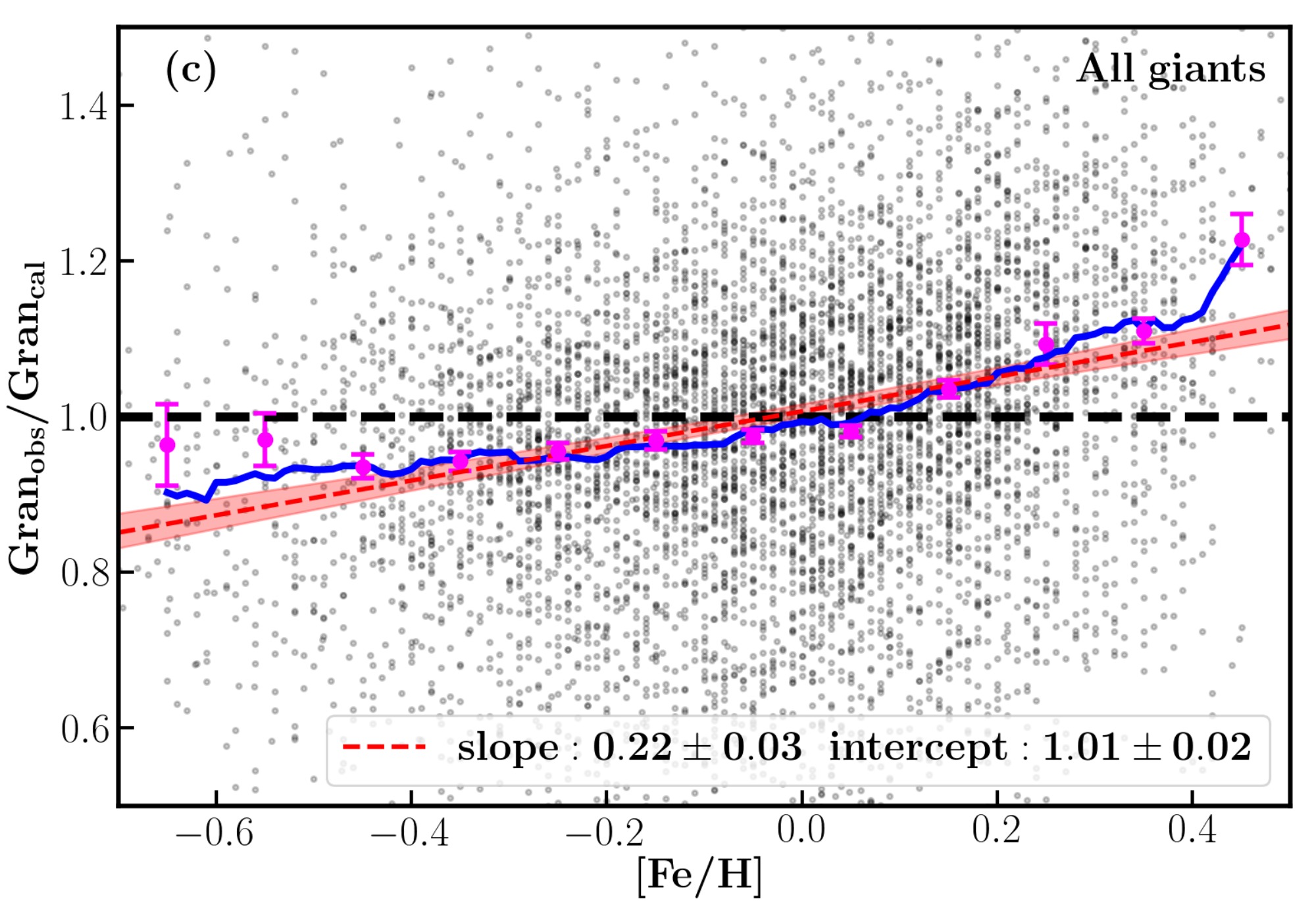}}\\
\caption{Same as Figure \ref{fig:fehresidual} but only using spectroscopic [Fe/H] from \citet{mathur17}.}
\label{fig:Fitfeh}
\end{center}
\end{figure}

To quantitatively measure the impact of metallicity on the oscillation amplitude and granulation power, we plot their residuals, 
in the sense of the observed divided by the fitted quantity, as a function of [Fe/H] in Figure \ref{fig:fehresidual}. The magenta points 
are the mean values in each [Fe/H] bin with a width of 0.1 dex. A linear fit is also shown with the red line. Clearly, 
the oscillation amplitude depends on metallicity, causing a 15\% variation across the metallicity range -0.7\textless[Fe/H]
\textless0.5 for both RGB and red clump stars, and so does granulation power, but with a 25\% scatter for the whole sample.    

Note that when investigating the metallicity influence, we used [Fe/H] estimates from \citet{mathur17}, of which 48\% originate 
from the KIC. \citet{pinsonneault14} found that the KIC metallicities show good agreement with those from APOGEE for red giant stars 
(see their Figure 14). We also performed the same analysis but only using spectroscopic metallicity from \citet{mathur17}, and 
found that the metallicity influence remains significant, as shown in Figure \ref{fig:Fitfeh}.

\section{Conclusions and Discussions}
We have presented a homogeneous analysis of 16094 oscillating red giants observed by \kep\ mission using all 
available end-of-mission long-cadence data sets. We provide a catalog of global seismic parameters and seismically derived 
mass, radius, and therefore surface gravity for oscillators, with \numax\ $>$ 5 \muHz. We have also systematically investigated 
the distribution of oscillation amplitude, granulation power, and width of power excess in RGB and HeB stars 
separately, and their dependencies on stellar mass and metallicity. The main results are summarized as follows:

\begin{itemize}
 \item We provide a catalog of seismic mass and radius and global oscillation parameters. The typical (median) uncertainties are  
      1.6\% for \numax, 0.6\% for \Dnu, 4.7\% for oscillation amplitude, 8.6\% for granulation power, 8.8\% for width of power excess, 
      7.8\% for mass, 2.9\% for radius, and 0.01 dex for \logg.  
       
\item We have improved the SYD pipeline to provide more accurate \Dnu\ estimates, some of which were incorrectly measured to be 
      $\Dnu \pm \delta \nu _{02}$. Our \numax\ and \Dnu\ measurements are in good agreement with the literature, displaying 
      a median fractional residual of 0.2\% and a scatter of 3.5\% for \numax, and a median fractional residual 
      of 0.01\% and a scatter of 4.2\% for \Dnu. 

\item We find that HeB stars form a extremely sharp edge, as shown in Fig \ref{dnunumax}b, which we interpret as the zero-age 
      main-sequence for core helium-burning. We also find tentative evidence for mass loss at the RGB tip and AGB phase.

\item RGB and HeB stars follow systematically different distributions of oscillation amplitude, power excess width, and granulation 
      power. Secondary clump stars have an overall lower oscillation amplitude and granulation power, and broader power excess than RGB 
      stars. This difference gradually attenuates toward lower-\numax\ RGB and HeB stars.

\item The oscillation amplitude and granulation power have dependencies on mass and metallicity. We confirm that the width of 
      power excess is an increasing function of mass. Metallicity has an influence on oscillation amplitude, leading to 15\% 
      variation for RGB stars and red clump stars in the metallicity range $-0.7<\rm{[Fe/H]}<0.5$, and on granulation power, causing a  
      25\% spread for all the red giants in the sample. \mbox{Metal-rich} and \mbox{lower-mass} stars show larger 
      oscillation amplitude and granulation power.   
\end{itemize}

Given the difficulty of appropriately fitting the power spectrum background, we do not report the measurement of oscillation 
amplitude,  width of power excess, or granulation power for stars with \numax\ $>$ 200 \muHz. We excluded all the stars 
with \numax\ $>$ 275 \muHz, though measurements of \Dnu\ using an autocorrelation function method are less affected \citep{yu16}. 

Recently, \citet{mathur17} delivered a characterization of the stellar fundamental properties of \kep\ targets for a transit 
detection run. This is based on conditioning stellar atmospheric parameters on the isochrones from the Dartmouth Stellar 
Evolution Database, which does not include helium-burning models for low-mass stars. Our seismic determinations of radius and 
mass are nearly independent of stellar models (except when correcting \Dnu) and therefore are able to remedy the bias of 
overestimated mass measurements for HeB stars.   
      
Our asteroseismic stellar properties can be used as reliable distance indicators and age proxies for mapping and dating 
the Galactic disk, as observed by the \kep\ telescope. It is also worthwhile to test and/or calibrate Gaia parallaxes. The 
precise and accurate seismically derived surface gravities lift the degeneracies from spectroscopically deriving atmospheric 
parameters. The $K2$ and $TESS$ missions are not expected to perform seismology on stars that are as distant and faint as this 
sample, so these stars will remain benchmark red giants for many years to come.

\citet{hon17} used our results for the successful classification of 5379 RGB and HeB 
stars. \citet{wu17} deduced the ages of the RGB stars in our sample and applied them as a training data set to derive ages and masses 
directly from LAMOST spectra. \citet{silvaaguirre17} chemically dissected the Milky Way disk population using a sample of 
red giant stars with the asteroseismic ages, which were determined using the global oscillation parameters measured from this work.

Our sample does not include stars oscillating with \numax\ below 5 \muHz. The scaling relations might not work appropriately 
because the low radial orders, \textit{n}, of observed modes cannot be reliably approximated with an asymptotic theory. Thus,  
the seismically inferred stellar masses and radii would be biased. However, it is of significant interest to measure the pulsation 
amplitudes and periods of those late RGB and AGB stars observed by \kep. This would be a great sample to investigate the 
correlation between the amplitude and period for long-period variables (LPVs, \citet{mosser13b, stello14}) and their relation 
with mass loss (J. Yu et al. in prep.).

\section*{Acknowledgments}
The authors would like to thank our anonymous referee for a careful reading of our manuscript and many helpful comments. We 
also want to thank Hans Kjeldsen for fruitful discussions. We gratefully acknowledge the entire \kep\ team and everyone 
involved in the \kep\ mission for making this paper possible. Funding for the \kep\ Mission is provided by NASA's Science Mission 
Directorate. Some/all of the data presented in this paper were obtained from the Mikulski Archive for Space Telescopes (MAST). 
STScI is operated by the Association of Universities for Research in Astronomy, Inc., under NASA contract NAS5-26555. Support 
for MAST for non-Hubble Space Telescope data is provided by the NASA Office of Space Science via grant NNX09AF08G and by other 
grants and contracts. D.H. acknowledges support by the Australian Research Council's Discovery Projects funding scheme (project 
number DE140101364) and support by the National Aeronautics and Space Administration under Grant NNX14AB92G issued through the 
Kepler Participating Scientist Program. D.S. is the recipient of an Australian Research Council Future Fellowship (project number 
FT1400147).

\bibliographystyle{aasjournal.bst}
\bibliography{/Users/jieyu/References/Literature/List/references}
\end{document}